%% file: main_revised2_black.tex
\documentclass[journal,10pt]{IEEEtran}
\renewcommand{\baselinestretch}{0.99}
\usepackage{hyperref}
\usepackage{cite}
\usepackage{amsmath,amsthm,amssymb,amsfonts,steinmetz}
\usepackage{graphicx,flushend}
\usepackage{xcolor}

\def\BibTeX{{\rm B\kern-.05em{\sc i\kern-.025em b}\kern-.08em
    T\kern-.1667em\lower.7ex\hbox{E}\kern-.125emX}}

\usepackage{tikz}
\usetikzlibrary{calc}
\makeatletter
\newcommand{\gettikzxy}[3]{%
  \tikz@scan@one@point\pgfutil@firstofone#1\relax
  \edef#2{\the\pgf@x}%
  \edef#3{\the\pgf@y}%
}
\makeatother

\usepackage{comment}
\usepackage{algorithm,algorithmic}
\usepackage{subcaption}

\input{./Definitions.tex}

\DeclareMathAlphabet\mathbfcal{OMS}{cmsy}{b}{n}

\begin{document}
\title{Electromagnetics-Compliant Optimization of\\ Dynamic Metasurface Antennas for Bistatic Sensing}

\author{Ioannis Gavras,~\IEEEmembership{Student~Member,~IEEE}, and George C. Alexandropoulos,~\IEEEmembership{Senior~Member,~IEEE}\vspace{-0.2cm} 
\thanks{Part of this work were recently presented at IEEE ICC, Montreal, Canada, June 2025~\cite{gavras2024circuit}.}
\thanks{This work has been supported by the SNS JU project TERRAMETA under the EU's Horizon Europe research and innovation programme under Grant Agreement number 101097101, including also top-up funding by UKRI under the UK government's Horizon Europe funding guarantee.}
\thanks{The authors are with the Department of Informatics and Telecommunications, National and Kapodistrian University of Athens, 16122 Athens, Greece. G. C. Alexandropoulos is also with the Department of Electrical and Computer Engineering, University of Illinois Chicago, IL 60601, USA (e-mails: \{giannisgav, alexandg\}@di.uoa.gr).}}

\maketitle
\begin{abstract}
Dynamic Metasurface Antennas (DMAs) are recently attracting considerable research interests due to their potential to enable low-cost, reconfigurable, and highly scalable antenna array architectures for next generation 
wireless systems. However, most of the existing literature relies on idealized models for the DMA operation, often overlooking critical structural and physical constraints inherent to their constituent metamaterials. In this paper, leveraging a recently proposed model incorporating physically consistent modeling of mutual coupling and waveguide propagation mapping, we optimize DMA transmission for bistatic sensing. A tractable approximation for the DMA response is first presented, which enables efficient optimization of the dynamically reconfigurable Lorentzian-constrained responses of the array's metamaterials. In particular, we formulate a robust beamforming optimization problem with the objective to minimize the worst-case position error bound in the presence of spatial uncertainties for the environment's scatterers as well as synchronization uncertainties at the multi-antenna receiver. To address the resulting high computational complexity due to the possibly excessive number of metamaterial-based antennas and their operation constraints, two low complexity beamforming designs are presented that perform offline searching over a novel beam codebook. In addition, capitalizing on the devised DMA optimization framework, a sequential parameter estimation
approach, performing coarse parameter acquisition followed by an alternating refinement stage, is presented. 
The validity of all proposed DMA beamforming designs in conjunction with the multi-target sensing scheme is assessed by means of Monte Carlo simulations for various system parameters, confirming that accurately modeling mutual coupling is essential for maintaining increased estimation performance. It is shown that, even under positioning and synchronization uncertainties, the proposed designs yield accuracy comparable to their fully digital and analog counterparts, while adhering to structural DMA constraints.  
\end{abstract}

\begin{IEEEkeywords}
Dynamic metasurface antennas, 
mutual coupling, OFDM, bistatic sensing, Cram\'{e}r-Rao bound, localization.
\end{IEEEkeywords}

\section{Introduction}

The advent of Sixth-Generation (6G) wireless networks heralds a new era of communications and sensing convergence, with unprecedented requirements in energy efficiency, localization accuracy, and sensing spatial resolution~\cite{8,6G-DISAC_mag}. As networks evolve towards higher frequencies (Frequency Ranges (FR) $2$ and $3$ \cite{shakya2024comprehensive}, and sub-THz), conventional antenna arrays face mounting challenges related to complexity, power consumption, and scalability \cite{THs_loc_survey, rappaport2017overview}. In response, metasurface-based antennas have emerged as a compelling alternative for realizing compact, reconfigurable, and cost-effective Radio Frequency (RF) front-ends for wireless transceivers~\cite{you2024next}. By leveraging sub-wavelength metamaterial elements with tunable Electromagnetic (EM) responses, programmable metasurfaces offer unparalleled flexibility in shaping and controlling wavefronts in real time, while consuming significantly less power compared to traditional phased arrays \cite{shaltout2019spatiotemporal,ramirez2025energy}. In this context, Dynamic Metasurface Antennas (DMAs) are lately gaining traction as an enabling antenna array technology, by combining metasurface reconfigurability with simplified hardware architecture, offering a practical and scalable solution to meet the stringent requirements of 6G systems \cite{20,castellanos2023energy}.

DMAs constitute a class of reconfigurable antenna panels composed of densely packed metamaterial elements with tunable impedance characteristics. These elements, typically organized along microstrips or rectangular waveguides, can be dynamically programmed to alter their phase and amplitude response to incident signals \cite{williams2022electromagnetic}. Unlike traditional phased arrays that require one phase shifter per antenna element and often depend on complex RF circuitry and numerous power-hungry active components, DMAs are capable of fine-grained analog BeamForming (BF) with a much smaller number of RF chains, even down to one per waveguide~\cite{zhang2022beam}. This design drastically reduces hardware complexity, power consumption, and cost, making DMAs highly attractive for wide-scale deployment in future Extremely Large (XL) Multiple-Input Multiple-Output (MIMO) systems~\cite{castellanos2023energy}. The core advantage of DMAs lies in their ability to manipulate wavefronts using programmable termination loads that follow Lorentzian dispersion models \cite{20}. These programmable loads, which are often realized using tunable capacitors, varactors, or switchable delay lines, enable sub-wavelength BF control without the need for individual signal generators or mixers per antenna element \cite{jabbar202460,dai2020reconfigurable}. Moreover, since DMA elements are deeply integrated into the EM substrate, they support extremely compact form factors and operate efficiently across a wide range of frequencies, including millimeter-Wave (mmWave) and sub-THz bands \cite{rasilainen2023hardware}. This makes them particularly well-suited for high-frequency applications, such as ultra-dense urban deployments, wearable devices, or aerial platforms, where traditional arrays would be infeasible due to size, cost, or power limitations.

State-of-the-art research on DMAs encompasses a broad spectrum of contributions, ranging from EM modeling and signal processing techniques to BF design and hardware implementations. For example, the authors in \cite{zhang2022beam} explored the use of DMAs in near-field multi-user MIMO systems, enabling beam focusing at precise spatial locations instead of just targeting specific directions, and proposed specialized algorithms to configure the Lorentzian-constrained analog weights for maximizing sum-rate performance. In \cite{Nlos_DMA}, online user tracking via a receiving DMA in non-line-of-sight, multipath-dominant environments and a scheme demonstrating high localization accuracy through ray-tracing-based simulations was devised, which outperformed previous emerging fingerprinting-based tracking methods. The authors in \cite{yang2023near} and \cite{gavras2025near} proposed iterative, codebook-based, and greedy optimization methods for configuring the Lorentzian-constrained weights of a receiving DMA for near-field localization in single- and multi-user scenarios, respectively. In both cases, the DMA-enabled systems achieved localization accuracy comparable to fully digital BF arrays, but with significantly reduced hardware complexity and cost. In\cite{gavriilidis2024metasurface} and \cite{gavras2024near}, the authors explored the communication and sensing capabilities, respectively, of DMA-based XL multi-antenna Receivers (RXs) under hardware constraints imposed by reception RF chains with $1$-bit Analog-to-Digital Converters (ADCs). In \cite{gavriilidis2024metasurface}, a closed-form expression for the achievable sum rate was derived and leveraged to design optimal analog and digital combiners that effectively mitigate the performance degradation caused by quantization. Meanwhile, \cite{gavras2024near} introduced a grid-search-based localization algorithm tailored for $1$-bit ADCs, achieving satisfactory localization accuracy comparable to systems without quantization limitations. Additionally, in \cite{NF_beam_tracking}, the authors studied the problem of near-field beam tracking in a high-frequency point-to-point wireless communication system between a DMA-equipped base station and a mobile single-antenna user. A theoretical analysis of the optimal achievable BF gain and a quantification of the performance loss due to user position mismatch were presented, addressing limitations of the relevant literature. 
In \cite{gavras2023full, gavras2024joint, alexandropoulos2025extremely, spawc2024}, full duplex XL MIMO transceiver architectures with DMAs were considered for joint communications and sensing. These works leveraged the partially-connected BF structure of DMAs to efficiently optimize their Lorentzian-constrained analog weights, enabling simultaneous communications and robust target localization in the challenging sub-THz frequency regime. It was showcased that DMA-based full duplex XL MIMO can match the performance of their fully digital BF counterparts, while reducing self-interference through a tailored low complexity BF design.

Despite the increasing research interests in DMAs, much of the existing open technical literature \cite{zhang2022beam,Nlos_DMA,37,yang2023near,gavras2025near,YXA2023,NF_beam_tracking,gavras2024joint,spawc2024,gavras2024simultaneous,alexandropoulos2025extremely,gavras2024near,gavras2025dma,gavras2023full,gavriilidis2024metasurface} relies on overly idealized models for their operation that overlook critical physical phenomena, potentially leading to misleading conclusions about the technology's true capabilities and limitations~\cite{gavriilidis2025microstrip}. In practical deployments, it is expected that dense arrangements of metamaterial elements within metasurface apertures will lead to strong mutual coupling, i.e., the excitation of one element will influence neighboring elements through both free-space radiation and guided-wave interactions along the shared waveguide~\cite{williams2022electromagnetic}. This effect, which becomes more pronounced at mmWave and sub-THz frequencies, introduces nonlinear and spatially varying distortions in the array response, fundamentally limiting the ability of DMA panels to replicate ideal beam patterns \cite{THs_loc_survey,dai2020reconfigurable}. Furthermore, waveguide-based signal propagation causes non-uniform power distribution and frequency-dependent attenuation, complicating the relationship between input RF signals and the resulting far-field radiation patterns \cite{williams2022electromagnetic,jabbar202460}. Consequently, for any practical DMA transceiver intended to be deployed for communications, localization, and/or sensing, the latter EM effects need to be explicitly accounted for to ensure reliable and robust operation. 

In this paper, we present a novel physically consistent framework for robust bistatic sensing of multiple targets relying on XL MIMO systems realized with the DMA technology. The adopted EM-compliant model for metasurface-based antenna panels captures mutual coupling effects, waveguide-based signal propagation, and Lorentzian-constrained metamaterial tuning, whereas our estimation framework accounts for target position uncertainties as well as synchronization inconsistencies at the RX side. To facilitate efficient designs of the DMA settings, a tractable approximation of the metasurface's response is first presented, which enables us to formulate and solve a robust localization optimization problem. 
The main contributions of the paper are summarized as follows.
\begin{itemize}
    \item Starting from the EM-compliant DMA model in~\cite{williams2022electromagnetic} that captures coupling effects between metamaterial elements, both through free-space interactions and inside waveguide propagation, we present a tractable approximation for the DMA response facilitating generic optimization of the metasurface's tunable parameters, while preserving high accuracy for various frequencies. 
    \item We formulate a novel robust sensing optimization problem minimizing the worst-case Position Error Bound (PEB) under uncertainties in the scatterers' positions and synchronization offsets at the bistatic system's RX side. The problem aims to jointly optimize the digital and analog BF weights at the Transmitter (TX), respecting the physical constraints of its DMA-based architecture.
    \item Recognizing the computational challenges of solving the proposed optimization problem formulation, we devise two low complexity solution alternatives: one based on a codebook-based strategy that constructs physically realizable beam patterns through projection onto the feasible DMA BF set, enabling efficient DMA-compatible codebook generation, and another based on an approximation of the original PEB-based objective via a lower bound.
    \item Starting from the likelihood function expressed with respect to the multi-target parameters of interest and nuisance parameters, an estimation approach that is consistent with the proposed PEB framework, performing sequential coarse parameters acquisition followed by respective alternating refinement stages, is presented.   
    \item Our extensive numerical results quantify the impact of mutual coupling and waveguide propagation mapping on the proposed DMA-based multi-target sensing performance. The interaction between target position estimation and TX-RX synchronization uncertainties under nonlinear distortions introduced by mutual coupling is demonstrated. Finally, it is showcased that the proposed EM-compliant robust sensing design offers comparable performance to the state-of-the-art physically ignorant benchmarks~\cite{gavras2025near, keskin2022optimal}.
\end{itemize}
 
A preliminary version of this paper's mutual-coupling-aware DMA framework was recently presented in~\cite{gavras2024circuit}, employing the EM-compliant DMA model from~\cite{williams2022electromagnetic}. However, the focus of this conference version was the optimization of a narrowband receiving DMA array for single target localization. In contrast, we herein address wideband DMA-enabled robust bistatic multi-target sensing, introducing a novel DMA response approximation that enables tractable optimization of~\cite{williams2022electromagnetic}'s EM-compliant DMA model. Furthermore, a novel codebook-based BF strategy compatible with DMA hardware is presented, which enables a tractable reformulation of the bistatic sensing design objective, which is then efficiently solved. Finally, unlike~\cite{gavras2024circuit} that solely evaluates sensing performance, this paper presents an in-depth investigation of the limitations of physically consistent DMA-based bistatic sensing under position uncertainties for the scatterers as well as TX-RX synchronization inconsistencies.

The remainder of this paper is organized as follows. Section~\ref{Sec: System} introduces the considered DMA-based bistatic sensing system together with the necessary models and the proposed tractable DMA response approximation for its thorough investigation. Section~\ref{Sec: Opt} introduces the proposed robust estimation optimization framework and describes the two designed low complexity DMA solution strategies. Section~\ref{Sec: num} presents our simulation setup and the detailed performance assessment of the proposed DMA designs, which are also compared  with state-of-the-art BF schemes and architectures. The concluding remarks of the paper, together with  our framework's future research directions, are included in Section~\ref{Sec: Conclusion}.

\textit{Notations:}
Vectors and matrices are represented by boldface lowercase and uppercase letters, respectively. The transpose, Hermitian transpose, inverse, and Moore-Penrose pseudoinverse of $\mathbf{A}$ are denoted as $\mathbf{A}^{\rm T}$, $\mathbf{A}^{\rm H}$, $\mathbf{A}^{-1}$, and $\mathbf{A}^{\rm \dagger}$, respectively. $\mathbf{I}_{n}$, $\mathbf{0}_{n}$, and $\boldsymbol{1}_n$ ($n\geq2$) indicate the $n\times n$ identity, zeros' matrices, and ones' column vector, respectively. $[\mathbf{A}]_{i,j}$ is the $(i,j)$-th element of $\mathbf{A}$, whereas notation $i:j$ as a matrix row/column index indicates its respective $i$-th till the $j$-th elements. $\|\mathbf{A}\|$ and $\|\mathbf{A}\|_{\rm F}$ represent $\mathbf{A}$'s Euclidean and Frobenious norms, respectively. $|a|$, ${\rm arg}(a)$, and $\Re\{a\}$ are respectively the amplitude, phase angle, and real part of complex scalar $a$, $\mathbb{C}$ is the complex number set, and $\jmath\triangleq\sqrt{-1}$ is the imaginary unit. $\mathbb{E}\{\cdot\}$ is the expectation operator and $\mathbf{x}\sim\mathcal{CN}(\mathbf{a},\mathbf{A})$ represents a complex Gaussian random vector with mean $\mathbf{a}$ and covariance matrix $\mathbf{A}$. Finally, ${\rm vec(\cdot)}$ and $\circ$ represent respectively the vectorization operation and the element-wise multiplication.

\begin{figure}[!t]
	\begin{center}
	\includegraphics[width=0.8\columnwidth]{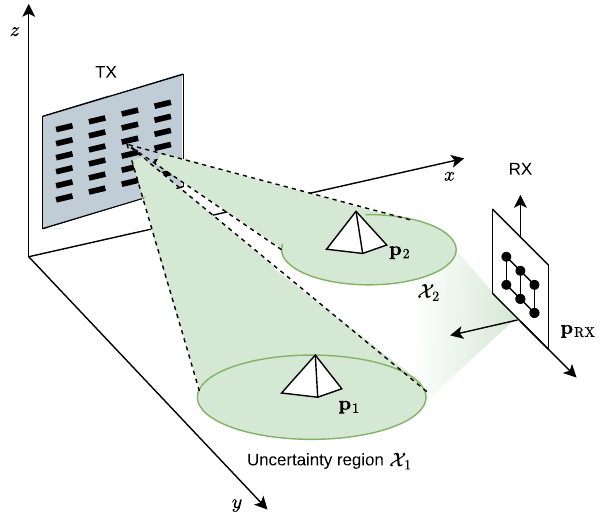}\vspace{-0.2cm}
	\caption{\small{The considered wideband bistatic sensing system comprising a DMA-equipped TX and a multi-antenna RX. Two out of the $G$ in total targets (modeled as scattering points) in the system's vicinity are illustrated, having the unknown exact positions $\mathbf{p}_1$ and $\mathbf{p}_2$ and the respective position uncertainty regions $\mathcal{X}_1$ and $\mathcal{X}_2$.}}\vspace{-0.4cm}
	\label{fig: DMA_System}
	\end{center}
\end{figure}

\section{System and DMA Response Models}\label{Sec: System}
We consider the wideband bistatic sensing system setup in Fig.~\ref{fig: DMA_System} consisting of a DMA-equipped TX and a multi-antenna RX wishing to localize $G$ targets lying in their vicinity, which are treated as Scattering Points (SPs).  
The TX's DMA is a Uniform Planar Array (UPA) comprising reconfigurable metamaterial elements with sub-wavelength spacing and tunable responses  (e.g., via semiconductor loading as in \cite{williams2022electromagnetic}), which are disjointly grouped into $N_{\rm RF}$ rectangular waveguides, with each waveguide connected to a dedicated transmit RF chain~\cite{williams2022electromagnetic, shlezinger2020dynamic}. The distance between adjacent waveguides is denoted by $d_{\rm RF}$. Each waveguide hosts $N_{\rm E}$ metamaterials arranged in a single dimension, with $d_{\rm E}$ being the distance between adjacent metamaterial elements. Hence, the total number of response-tunable metamaterials at the TX DMA is $N_{\rm T} \triangleq N_{\rm RF}N_{\rm E}$. On the other side of the sensing system, the multi-antenna RX consists of $M_{\rm R}$ reception RF chains, each connected to a distinct antenna element, thus, RX possesses in total $M_{\rm R}$ antennas. This enables it to realize fully digital combining of the signals received at its antennas, which carry spatial information about the $G$ targets\footnote{In this paper, we assume that more than single-bounce reflections of the TX signals from the $G$ targets lying in the system's vicinity are highly attenuated, and can be thus neglected \cite{christopher2019single}.}. This BF operation is modeled by the matrix $\mathbf{W}_{\rm RX} \in \mathbb{C}^{M_{\rm R} \times M_{\rm R}}$.

The considered wideband bistatic sensing system relies on Orthogonal Frequency-Division Multiplexing (OFDM) pilots. In particular, the TX transmits $T$ OFDM pilots per frame over $K$ SubCarriers (SCs), which are repeated over $M$ consecutive frames. To ensure the feasibility of the BF design for sensing, we assume that $G<{\rm \min}\{N_{\rm RF},M_{\rm R}\}$. The pilot symbol matrix for each $m$-th frame ($m=1,\ldots,M$) is denoted by $\S_m\in\Compl^{K\times T}$,  where the element $s_{m,k,t}\triangleq\left[\S_m\right]_{k,t}$ $\forall k,t$, with $|s_{m,k,t}|^2=1$, represents the pilot symbol transmitted on the $k$-th SC during the $t$-th transmission. We assume that each symbol within each the $m$-th pilot frame is firstly digitally processed via the digital BF vector $\f_m\in\Compl^{N_{\rm RF}\times 1}$ and then analog processed using the analog BF matrix $\W_{\rm TX}\in\Compl^{N_{\rm T}\times N_{\rm RF}}$; the extension to the more complex frequency selective hybrid BF case is left for future work. To this end, the resulting $N_{\rm T}$-element transmitted signal is represented as $\x_{m,k,t}\triangleq\W_{\rm TX}\f_ms_{m,k,t}$, which is assumed to be power limited such that $\mathbb{E}\{\|\W_{\rm TX}\F\|_{\rm F}^2\}\leq P_{\rm tot}/M$, where $\F\triangleq[\f_1,\ldots,\f_M]\in\Compl^{N_{\rm RF}\times M}$ and  $P_{\rm tot}$ represents the total TX power budget across all $M$ frames. It is noted that the matrix $\mathbf{W}_{\rm TX}$ models the DMA-based mapping between the digitally processed pilot symbols and those actually transmitted over the air through DMA manipulation, which will be next detailed. 

\subsection{Received Signal and Channel Models}
Assuming that the bistatic sensing system operates in the far-field region, the $M_{\rm R}$-element received signal vector at the baseband of the RX side on the $k$-th SC during the $t$-th pilot symbol transmission of the $m$-th frame can be mathematically expressed as follows:
\begin{align}\label{eq: rec_signal}
    \y_{m,k,t} = \W_{\rm RX}^{\rm H}\H_k\W_{\rm TX}\f_ms_{m,k,t}+\n_{m,k,t},
\end{align}
where $\n_{m,k,t}\sim\mathcal{CN}(\mathbf{0}_{M_{\rm R}\times 1},\sigma^2\mathbf{I}_{M_{\rm R}})$ represents the corresponding Additive White Gaussian Noise (AWGN) vector, and $\H_k\in\Compl^{M_{\rm R}\times N_{\rm T}}$ is the end-to-end composite channel matrix at the $k$-th SC, which is modeled as follows:
\begin{align}
    \H_k = \sum_{g=1}^G a_g e^{-\jmath2\pi f_k\tau_g}\a_{\rm RX}\left(\psi_{{\rm a},g},\psi_{{\rm e},g}\right)\a_{\rm TX}^{\rm H}\left(\theta_{{\rm a},g},\theta_{{\rm e},g}\right).
\end{align}
In this expression, $a_g$ is the complex path gain, $\tau_g$ is the time delay, and $\theta_{{\rm a},g}$, $\theta_{{\rm e},g}$ are the azimuth and elevation Angles of Departure (AoDs), while $\psi_{{\rm a},g}$, $\psi_{{\rm e},g}$ are the corresponding azimuth and elevation Angles of Arrival (AoAs). The steering vectors at the TX DMA and the RX are denoted by $\a_{\rm TX}(\cdot) \in \Compl^{N_{\rm T} \times 1}$ and $\a_{\rm RX}(\cdot) \in \Compl^{M_{\rm R} \times 1}$, respectively. Note that both of these nodes in our bistatic sensing system setup are assumed to have complete knowledge of each other's 3D positions. This implies that the RX is capable of effectively canceling its Line-of-Sight (LoS) path with the TX~\cite{saini2003direct}, however, the former node remains unsynchronized relative to the latter.

As shown in the 3D Cartesian coordinate system in Fig.~\ref{fig: DMA_System}, the TX is positioned at the origin, while the RX is located at the point $\p_{\rm RX}\triangleq[x_{\rm RX},y_{\rm RX},z_{\rm RX}]$, facing the TX with an orientation angle $\zeta \in [-0.5\pi,0.5\pi]$ (i.e., the angular range of rotation within which localization remains feasible), which  represents its rotation around the $z$-axis. Due to the far-field propagation assumption, each SP is modeled as a point source positioned between the TX and RX at the points $\p_g\triangleq[x_g,y_g,z_g]$ $\forall g=1,\ldots,G$. These positions constitute the estimation objective of the considered bistatic sensing system, and we make the assumption that each $\p_g$ is linked with a spherical position uncertainty region defined as follows: 
\begin{equation}
\mathcal{X}_g = \{\x \in \mathbb{R}^3 \mid \|\x - \p_g\| \leq u_g\},
\end{equation}

where $u_g$ denotes the radius determining the uncertainty bound. It is assumed that $\mathcal{X}_g$ $\forall g$ are available to the bistatic system via a dedicated tracking process~\cite{mendrzik2019enabling}, e.g.: \textit{i}) tracking across sensing epochs (e.g., Kalman/particle filtering) providing a predicted location and covariance \cite{BKC2017}; \textit{ii}) coarse acquisition (e.g., wide-beam scans/codebook sweeps) defining a spatial gate \cite{FCWC2022}; \textit{iii}) external information (e.g., GNSS or network-aided positioning) \cite{CYH25,MGKW2017}; or \textit{iv}) situational-awareness procedures~\cite{mendrzik2019enabling}. The radius \(u_g\) captures the quality of this prior information: larger \(u_g\) implies greater uncertainty inducing a more conservative design that trades peak gain for coverage, with the region center and spatial extent determined by the corresponding mean and covariance.

In addition, the TX steering vector (with a similar definition applying to RX's steering vector) with respect to each $g$-th path is given by $\a_{\rm TX}(\theta_{{\rm a},g},\theta_{{\rm e},g})\triangleq\a_{\rm x}(\theta_{{\rm a},g},\theta_{{\rm e},g})\otimes\a_{\rm y}(\theta_{{\rm a},g},\theta_{{\rm e},g})$, where:
\begin{align}
\nonumber\a_{\rm x}(\theta_{{\rm a},g},\!\theta_{{\rm e},g})&\!=\!\frac{1}{\sqrt{N_{\rm E}}}[1,\ldots,e^{\jmath\frac{2\pi}{\lambda}d_{\rm E}(N_{\rm E}-1)\!\sin\theta_{{\rm e},g}\!\cos\theta_{{\rm a},g}}]^{\rm T},\\
\nonumber\a_{\rm y}(\theta_{{\rm a},g},\!\theta_{{\rm e},g})&\!=\!\frac{1}{\sqrt{N_{\rm RF}}}[1,\ldots,e^{\jmath\frac{2\pi}{\lambda}d_y(N_{\rm RF}-1)\!\sin\theta_{{\rm e},g}\!\sin\theta_{{\rm a},g}}]^{\rm T}.
\end{align}
According to this system geometry, the involved AoDs and AoAs in~\eqref{eq: rec_signal} can be expressed as follows:
\begin{align}
    &\nonumber\psi_{{\rm a},g} = {\rm atan2}(y_g-y_{\rm RX},x_g-x_{\rm RX}),\theta_{{\rm e},g} =\arcsin{\left(\frac{z_g}{\|\p_g\|}\right)},
    \\&\nonumber \psi_{{\rm e},g}=\arcsin{\left(\frac{z_g-z_{\rm RX}}{\|\p_g-\p_{\rm RX}\|}\right)},\theta_{{\rm a},g} = {\rm atan2}(y_g,x_g)-\zeta,
\end{align}
with ${\rm atan2}(\cdot)$ representing the four-quadrant inverse tangent. To this end, the corresponding time delays and channel gains are given by the following expressions:
\begin{align}
    &\nonumber\tau_{g}=\left(\|\p_g\|+\|\p_{\rm RX}-\p_g\|\right)/c+\Delta t,\\
    &\nonumber a_{g} = \beta_g\lambda F(\theta_{{\rm e},g})/\left(4\pi\left(\|\p_g\|+\|\p_{\rm RX}-\p_g\|\right)\right),
\end{align}
where $\Delta t$ is the clock offset between the TX and the unsynchronized RX, modeled as $\Delta t\sim\mathcal{N}(0,\sigma^2_{\rm clk})$, $\beta_g$ denotes the $g$-th SP's complex-valued reflection coefficient; and $F(\cdot)$ represents each metamaterial's radiation profile, which is modeled for an elevation angle $\theta_{\rm e}$ as follows:
\begin{align*}
    F (\theta_{\rm e}) = \begin{cases}
    2(b_{\rm bor}+1)\cos^{b_{\rm bor}}(\theta_{\rm e}),& {\rm if}\, \theta_{\rm e}\in[-0.5\pi,0.5\pi]\\
    0,              & {\rm otherwise}
\end{cases}.
\end{align*}
In the latter expression, $b_{\rm bor}$ determines the boresight antenna gain which depends on the specific DMA technology \cite{ellingson2021path}.

\subsection{DMA Response Model}\label{Sec: MC_DMA}
Capitalizing on~\cite{williams2022electromagnetic}, the analog BF matrix $\W_{\rm TX}$ accounts for wave propagation, reflections within the waveguides feeding the transmit RF chains, and mutual coupling effects, both through the air and within the waveguides. To this end, it is mathematically expressed as follows:
\begin{align}\label{eq: BF}
    \W_{\rm TX} = (\Q_{\rm TA}+\W_{\rm MC})^{-1}\P_{\rm SA},
\end{align}
where $\P_{\rm SA} \in \Compl^{N_{\rm T} \times N_{\rm RF}}$ represents the signal propagation from each RF chain to its associated metamaterial element within their respective waveguide. Consequently, its entries are zero for element pairs in different waveguides, as they lack a shared RF chain. In particular, the entries of $\P_{\rm SA}$ are given $\forall$$n=1,\ldots,N_{\rm T}$ and $\forall$$i=1,\ldots,N_{\rm RF}$ as follows:
\begin{align}
    \nonumber&[\P_{\rm SA}]_{n,i} \\ \nonumber&=\begin{cases}
    \jmath\omega\epsilon{G}_{\rm SA}(\p_n,\p_i),&  \text{for $\p_n,\p_i$ in the same waveguides}\\
    0,              & \text{for $\p_n,\p_i$ in different waveguides}
\end{cases},
\end{align}
where $\p_n\triangleq [x_n, y_n, z_n]$ and $\p_i\triangleq [x_i, y_i, z_i]$ denote the $3$-Dimensional (3D) cartesian coordinates of each $n$-th element connected to each $i$-th transmit RF chain and the respective output port, respectively. Additionally, ${G}_{\rm SA}(\p_n,\p_i)$ represents the third diagonal component of the Green’s function inside the waveguide~\cite{williams2022electromagnetic}, which captures the waveguide propagation between any pair of points $\p_n$ and $\p_i$ and is given as:
\begin{align}\label{eq: Green}
     &G_{\rm SA}\left(\p_n, \p_i\right)\triangleq\frac{-k_x \sin \left(\frac{\pi z_n}{a}\right) \sin \left(\frac{\pi z_i}{a}\right)}{a b k^2 \sin \left(k_x S_\mu\right)} \\\nonumber&\times\left[\cos \left(k_x\left(x_i+x_n-S_\mu\right)\right)+\cos \left(k_x\left(S_\mu-\left|x_n-x_i\right|\right)\right)\right],
\end{align}
where $a, b$, and $S_{\mu}$ denote the waveguide’s width, height, and length, respectively. Furthermore, $k_x$ is defined as $k_x\triangleq\Re\{\sqrt{k^2-(\pi/a)^2}\}-\jmath\Im\{\sqrt{k^2-(\pi/a)^2}\}$, where $k$ is the waveguide’s wavenumber that is given by $k\triangleq2\pi\lambda^{-1}\sqrt{\epsilon_r\mu_r}$ with $\epsilon_r$ and $\mu_r$ being respectively the relative permittivity and relative permeability\footnote{The Green’s function for the waveguide propagation given by~\eqref{eq: Green} is evaluated under the single-mode rectangular waveguide assumption~\cite{williams2022electromagnetic}, i.e., only the dominant transverse electric mode (denoted by $\mathrm{TE}_{10}$ in~\cite{williams2022electromagnetic}) is considered. Accordingly, the wavenumber is $k_x=\sqrt{k^2-(\pi/a)^2}$ and $\mathrm{TE}_{10}$'s cutoff frequency is $f_{c,10}=\frac{c}{2a\sqrt{\varepsilon_r\mu_r}}$. This implies that, for $f_c>f_{c,10}$, the mode propagates (i.e., $k_x\in\mathbb{R}$), whereas, for $f\leq f_{c,10}$, it becomes evanescent (i.e., $k_x$ is purely imaginary)~\cite{ChewWaveguidesNotes}. Since dispersion increases as $f_c$ approaches $f_{c,10}$, the DMA response becomes increasingly frequency selective near cutoff. In this paper, following~\cite{williams2022electromagnetic}, we assume operation sufficiently above $f_{c,10}$ and below higher mode cutoffs, where the array response varies only mildly across the occupied bandwidth. For this case, we adopt a DMA approximation that is frequency flat, and apply the same analog BF matrix $\W_{\rm TX}\in\Compl^{N_{\rm T}\times N_{\rm RF}}$ given by~\eqref{eq: BF} to all $K$ SCs.}, and $\lambda\triangleq c/f_c$ represents the wavelength corresponding to the operating frequency $f_c$, with notation $c$ representing the speed of light.

A key challenge in the dense DMA architecture is the mutual coupling between adjacent radiating elements due to their sub-wavelength spacing. This coupling occurs both through the air between surface elements and within the waveguide for elements in the same waveguide.

To this end, each $(n,n')$-th entry of $\W_{\rm MC} \in \Compl^{N_{\rm T} \times N_{\rm T}}$ $\forall$$n,n'=1,2,\ldots,N_{\rm T}$ can be calculated for each $(\p_n,\p_{n'})$ pair in the same waveguide (with $\p_n$ and $\p_{n'}$ being the 3D cartesian coordinates for each $n$-th and $n'$-th metamaterial, respectively) as follows~\cite{williams2022electromagnetic}:
\begin{align}
    (\jmath\omega\epsilon)^{-1}[\W_{\rm MC}]_{n,n'} = 2G_{\rm MC}(\p_n,\p_{n'})+G_{\rm SA}(\p_n,\p_{n'}),
\end{align}
whereas, for $\p_n$ and $\p_{n'}$ placed in different waveguides, $(\jmath\omega\epsilon)^{-1}[\W_{\rm MC}]_{n,n'}=2G_{\rm MC}(\p_n,\p_{n'})$. In the latter expression, $\epsilon$ denotes the medium’s permittivity and $G_{\rm MC}(\p_n,\p_{n'})$ represents the third diagonal component of the Green's function in free space, which is given by:
\begin{align}
     &G_{\rm MC}(\p_n,\p_{n'})  \\\nonumber&=\left(\frac{R^2-\Delta z^2}{R^2}-\jmath\frac{R^2-3\Delta z^2}{R^3k}\right)\left(\frac{R^2-3\Delta z^2}{R^4k^2}\right)\frac{e^{-\jmath kR}}{4\pi R},
\end{align}
where $R \triangleq \|\p_n-\p_{n'}\|$ and $\Delta z\triangleq z_n-z_{n'}$. 
Note that the above expressions apply only to distinct element pairs
$n\neq n'$ located within the same waveguide. For $n=n'$, the
free-space Green's function becomes singular, and the corresponding
diagonal entry of $\W_{\rm MC}$ is instead defined as the regularized
self-admittance of the $n$-th metamaterial element within that
waveguide. Following~\cite{williams2022electromagnetic}, this quantity
is obtained by taking the co-location limit
$\p_{n'}\rightarrow\p_n$, which, for the considered finite-length
rectangular-waveguide model, yields the following closed-form
expression:
\begin{align}\nonumber
[\W_{\rm MC}]_{n,n}
=
\frac{k\omega\epsilon}{3\pi}
\!-\!
\jmath
\frac{
k_x\left[
\cos\!\left(k_x(S_\mu-2x_n)\right)
-
\cos\!\left(k_xS_\mu\right)
\right]
}{
ab\omega\mu\sin\!\left(k_xS_\mu\right)
},
\end{align}
where $\mu\triangleq\mu_0\mu_r$ denotes the magnetic permeability of the waveguide medium, with $\mu_0$ representing the vacuum permeability~\cite{williams2022electromagnetic}.

Last but not least, $\mathbf{Q}_{\rm TA}\in\Compl^{N_{\rm T}\times N_{\rm T}}$ in~\eqref{eq: BF} is a diagonal matrix collecting the tunable admittance values associated with the metamaterial terminations. Following the EM-compliant admittance formulation in~\cite{williams2022electromagnetic} for passive elements, lossless reactive tuning of each metamaterial termination is assumed. In particular, the corresponding diagonal entry of matrix $\mathbf{Q}_{\rm TA}$ is defined as follows:
\begin{align}
\left[\mathbf{Q}_{\rm TA}\right]_{n,n}
\triangleq
\jmath\left(c_n-\Im\left\{\left[\mathbf{W}_{\rm MC}\right]_{n,n}\right\}\right),
\label{eq:QTA_reactive_tuning}
\end{align}
where $c_n\in\mathbb{R}$ denotes the tunable reactive part. The total admittance entering the element response is given by the sum between $\mathbf{Q}_{\rm TA}$ and a diagonal matrix collecting the diagonal entries of the mutual coupling matrix $\mathbf{W}_{\rm MC}$. Therefore, for each $n$-th metamaterial element, we obtain $\left[\mathbf{Q}_{\rm TA}+\mathbf{W}_{\rm MC}\right]_{n,n}=R_n+\jmath c_n,$ where $c_n=-R_n\tan\left(\frac{\varphi_n}{2}\right)$ and $R_n\triangleq\Re\{[\mathbf{W}_{\rm MC}]_{n,n}\}>0$ denotes the positive real part of the admittance entering the element response. Equivalently, by introducing the phase variable $\varphi_n\in(-\pi,\pi)$, the total admissible admittance can be parameterized as follows:
\begin{align}
\left[\mathbf{Q}_{\rm TA}+\mathbf{W}_{\rm MC}\right]_{n,n}
=
R_n\left(
1-\jmath\tan\left(\frac{\varphi_n}{2}\right)
\right).
\label{eq:QTA_lorentzian_admittance}
\end{align}
Under this parametrization, the corresponding scalar element response is formulated as:
\begin{align}
\vartheta_n
\triangleq
\left(
\left[\mathbf{Q}_{\rm TA}+\mathbf{W}_{\rm MC}\right]_{n,n}
\right)^{-1}
=
\frac{1}{2R_n}
\left(1+e^{\jmath\varphi_n}\right),
\label{eq:lorentzian_response}
\end{align}
which is the Lorentzian amplitude-phase constrained response that is commonly used~\cite{williams2022electromagnetic}.

\subsubsection{Approximation for $\W_{\rm TX}$}\label{Sec: MC}
The vast majority of the DMA-based literature~\cite{Nlos_DMA,37,yang2023near,gavras2025near,YXA2023,NF_beam_tracking,gavras2024joint,spawc2024,gavras2024simultaneous,alexandropoulos2025extremely,gavras2024near,gavras2023full,gavriilidis2024metasurface} adopts a simplified version of~\eqref{eq: BF} neglecting mutual coupling, thus, allowing $\mathbf{Q}_{\rm TA}$ included in this expression to be readily optimized for various objectives. In contrast, the physically-consistent model in~\eqref{eq: BF} poses significant challenges when considered for optimization, due to the presence of the inverse of a sum involving the termination admittance and mutual coupling matrices. To address this, we adopt an approximation approach inspired by~\cite{FSA2023,RLS2024}, enabling more tractable system design and optimization. Specifically, we apply the Woodbury matrix identity with respect to $\Q_{\rm TA}$ to expand the matrix inverse in~\eqref{eq: BF} as follows:
\begin{align}
    \left(\Q_{\rm TA}\!+\!\W_{\rm MC}\right)^{\rm -1} &= \Q_{\rm TA}^{\rm -1}\!-\!\Q_{\rm TA}^{\rm -1}\left(\Q_{\rm TA}\W_{\rm MC}^{\rm -1}\!+\!\I_N\right)^{\rm -1}\!.
\end{align}
Next, since the condition $\|\Q_{\rm TA}\W_{\rm MC}^{-1}\|_{\rm F}^2\ll1$ is almost surely satisfied\footnote{This condition is generally satisfied in practice for dense antenna arrays such as DMAs, where the mutual coupling matrix $\W_{\rm MC}$ captures strong inter-element interactions, leading to a large-norm matrix. Consequently, $\W_{\rm MC}^{-1}$ has a small norm, and since $\Q_{\rm TA}$ is a diagonal matrix with Lorentzian-constrained entries, its spectral norm is bounded. This ensures that the product $\Q_{\rm TA}\W_{\rm MC}^{-1}$ remains small in terms of its norm value, satisfying the convergence condition of the Neumann expansion~\cite{RLS2024}.}, we can apply the Neumann series expansion to the term $\left(\Q_{\rm TA}\W_{\rm MC}^{\rm -1}+\I_N\right)^{\rm -1}$, yielding the approximation:
\begin{align}
    \left(\Q_{\rm TA}\W_{\rm MC}^{\rm -1}\!+\!\I_N\right)^{\rm -1} \approx \I_N\!+\!\sum_{p=1}^{\infty}(-1)^p\left(\Q_{\rm TA}\W_{\rm MC}^{\rm -1}\right)^p\!.
\end{align}
Combining the previously derived expressions results in:
\begin{align}\label{eq: dma_approx}
    \left(\Q_{\rm TA}\!+\!\W_{\rm MC}\right)^{\rm -1}\approx -\Q_{\rm TA}^{-1}\sum_{p=1}^{\infty}(-1)^p\left(\Q_{\rm TA}\W_{\rm MC}^{\rm -1}\right)^p\!.
\end{align}

\begin{figure}[!t]
	\begin{center}
	\includegraphics[width=\columnwidth]{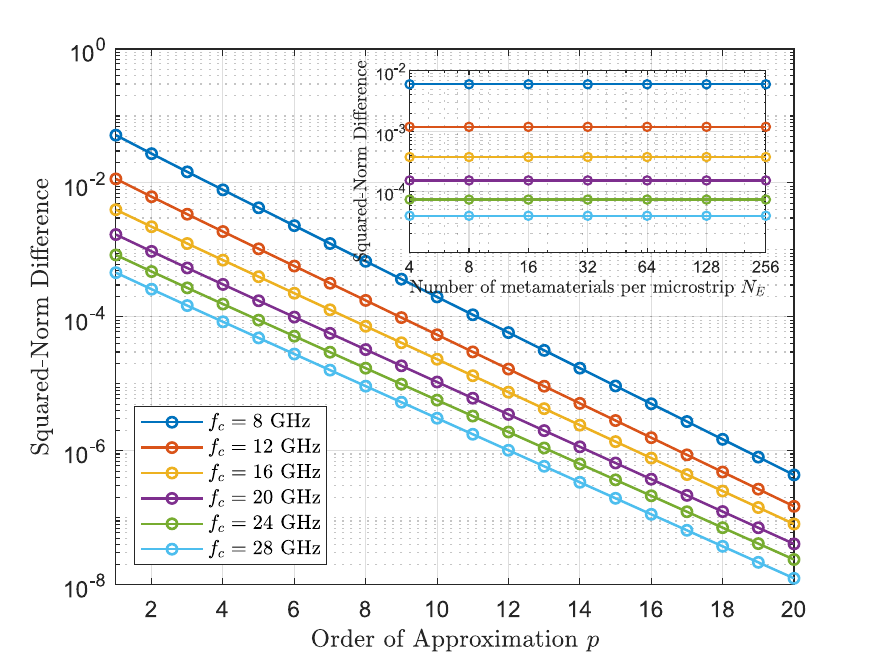}
	\caption{\small{Squared-norm difference between the exact and approximated values of $(\Q_{\rm TA}+\W_{\rm MC})^{-1}$ versus the approximation order $p$ for various carrier frequencies $f_c$, using a TX DMA panel with $N_{\rm RF} = 4$ waveguides, each containing $N_{\rm E} = 32$ metamaterial elements, and $(d_{\rm RF},d_{\rm E})=(\lambda/2,\lambda/5)$. The inset figure shows the squared-norm difference versus the number of metamaterials $N_{\rm E}$ per rectangular waveguide.}}
    \vspace{-0.4cm}
	\label{fig: approx}
	\end{center}
\end{figure}
Let us replace~\eqref{eq: dma_approx} using only its first two terms (i.e., for $p=1$ and $2$) in~\eqref{eq: BF}. This yields the following second-order approximation of the mutual-coupling-aware DMA response model that provides an effective simplification of the analog BF matrix at the TX side of the bistatic sensing system:
\begin{align}\label{eq: sec_order}
    \W_{\rm TX}&\approx\left(\W_{\rm MC}^{-1}-\W_{\rm MC}^{-1}\Q_{\rm TA}\W_{\rm MC}^{-1}\right)\P_{\rm SA}\nonumber\\
    &=\left(\W_{\rm MC}^{-1}-\W_{\rm MC}^{-1}\jmath{\rm diag}(\c-\boldsymbol{\nu})\W_{\rm MC}^{-1}\right)\P_{\rm SA}.
\end{align}
where $\boldsymbol{\nu}\triangleq\Im\left\{\operatorname{diag}(\W_{\rm MC})\right\}\in\mathbb{R}^{N_{\rm T}\times 1}$ and $\c\triangleq[c_1,\ldots,c_{N_{\rm T}}]^{\rm T}\in\mathbb{R}^{N_{\rm T}\times 1}$ collects the tunable reactive weights. In Fig.~\ref{fig: approx}, we investigate the accuracy of the $\W_{\rm TX}$ approximation through~\eqref{eq: dma_approx}, including the latter second-order one in~\eqref{eq: sec_order}. In particular, this figure illustrates the squared-norm difference between the true and approximated values of $\left(\Q_{\rm TA} + \W_{\rm MC}\right)^{-1}$ versus the approximation order $p$ across different operating frequencies $f_c$ within the FR3 band (approximately $7$-$24$ GHz \cite{shakya2024comprehensive}), considering a TX DMA with $N_{\rm RF} = 4$ waveguides, $(d_{\rm RF},d_{\rm E})=(\lambda/2,\lambda/5)$, and various values for the number $N_{\rm E}$ of metamaterial elements per rectangular waveguide. As shown in the figure, the squared-norm difference decreases nearly exponentially with the approximation order $p$ for all carrier frequencies; lower errors are demonstrated at higher frequencies. As observed, while higher-order terms further improve accuracy, the second-order approximation offers a favorable accuracy--complexity tradeoff; higher-order terms involve repeated products with $\Q_{\rm TA}$ and quickly become cumbersome. The inset further shows that, for the considered fixed spacing, the error is nearly insensitive to increasing $N_{\rm E}$, and the same trend persists for larger $N_{\rm RF}$.

\section{EM-Compliant DMA Design for Sensing}\label{Sec: Opt}
In this section, we first derive the PEB metric quantifying the multi-target localization capability of the proposed DMA-based bistatic sensing system setup. Next, capitalizing on the physically-consistent DMA response model in Section~\ref{Sec: MC}, we present a robust optimization framework for the design of the TX DMA's reconfigurable parameters. 

\subsection{Cram\'{e}r-Rao Bound (CRB) Derivation}\label{sec: CRB}
Let us introduce $\boldsymbol{\xi}\triangleq[\boldsymbol{\theta}_{\rm a}^{\rm T},\boldsymbol{\theta}_{\rm e}^{\rm T},\boldsymbol{\psi}_{\rm a}^{\rm T},\boldsymbol{\psi}_{\rm e}^{\rm T},\boldsymbol{\alpha}_R^{\rm T},\boldsymbol{\alpha}_I^{\rm T},\boldsymbol{\tau}^{\rm T}]^{\rm T}\in\mathbb{R}^{7G\times 1}
$ and $\widetilde{\boldsymbol{\xi}}\triangleq[\p^{\rm T},\zeta,\Delta t]^{\rm T}\in\mathbb{R}^{(3G+2)\times 1}$ including the unknown channel and location parameters, respectively, where $\boldsymbol{\theta}_{\rm a}\triangleq[\theta_{{\rm a},1},\ldots,\theta_{{\rm a},G}]$, $\boldsymbol{\theta}_{\rm e}\triangleq[\theta_{{\rm e},1},\ldots,\theta_{{\rm e},G}]$, $\boldsymbol{\psi}_{\rm a}\triangleq[\psi_{{\rm a},1},\ldots,\psi_{{\rm a},G}]$, $\boldsymbol{\psi}_{\rm e}\triangleq[\psi_{{\rm e},1},\ldots,\psi_{{\rm e},G}]$, $\boldsymbol{\alpha}_R\triangleq[\Re\{a_1\},\ldots,\Re\{a_G\}]$, $\boldsymbol{\alpha}_I\triangleq[\Im\{a_1\},\ldots,\Im\{a_G\}]$, $\boldsymbol{\tau}\triangleq[\tau_1,\ldots,\tau_G]$, and $\p\triangleq[\p_1,\ldots,\p_G]^{\rm T}\in\mathbb{R}^{3G\times 1}$. It is evident from \eqref{eq: rec_signal}'s inspection that, within each $m$-th frame during each $t$-th transmission and over each $k$-th SC, the received signal at the outputs of the RX's reception RF chains can be modeled as $\y_{m,k,t}\sim\mathcal{CN}(\boldsymbol{\mu}_{m,k,t},\sigma^2\I_{M_{\rm R}})$ with $\boldsymbol{\mu}_{m,k,t} \triangleq \W_{\rm RX}^{\rm H}\H_k\W_{\rm TX}\f_ms_{m,k,t}$. Each $(i,j)$-th element (with $i,j=1,\ldots,7G$) of the Fisher Information Matrix (FIM) $\J \in \Compl^{7G\times7G}$ associated with the channel parameters' vector $\boldsymbol{\xi}$ can be computed as follows \cite{kay1993fundamentals}:
\begin{align}\label{eq: FIM}
    [\J]_{i,j}\!=\! \frac{2}{\sigma^2}\sum_{m=1}^M\sum_{k=1}^K\sum_{t=1}^T\Re\left\{\!\frac{\partial \boldsymbol{\mu}^{\rm H}}{\partial[\boldsymbol{\xi}]_i}\frac{\partial \boldsymbol{\mu}}{\partial[\boldsymbol{\xi}]_j}\!\right\}.
\end{align}
Focusing on the SPs' position estimation, we next deploy the
transformation matrix $\T\in\mathbb{R}^{7G\times (3G+2)}$, which can be expressed as a Jacobian with $[\T]_{i,j}=\partial[{\boldsymbol{\xi}}]_i/\partial[\widetilde{\boldsymbol{\xi}}]_j$ $\forall i,j$, to derive the FIM $\widetilde{\J}\in\Compl^{(3G+2)\times(3G+2)}$ of the locations parameters' vector $\widetilde{\boldsymbol{\xi}}$, as follows:
\begin{align}\label{eq: EFIM}
    \widetilde{\J} = \T^{\rm T}\J\T+\J_{\rm prior},
\end{align}
where $\T^{\rm T}\J\T$ and $\J_{\rm prior}$ represent the FIMs with respect to the observations and the available prior knowledge, respectively. Since only the clock bias is assumed to be random in $\widetilde{\boldsymbol{\xi}}$, it holds that $[\J_{\rm prior}]_{3G+2,3G+2}=\sigma_{\rm clk}^{-2}$ and all its other entries will be zero~\cite{keskin2022optimal}. Recall that the RX's position is not included in $\widetilde{\boldsymbol{\xi}}$, since it is assumed to be known at the TX, and vice versa. 

To assess the accuracy of the SP position estimates, we use the PEB as the performance metric, which is computed as:
\begin{align}\label{eq: PEB}
    {\rm PEB}\left(\Q_{\rm TA},\F;\widetilde{\boldsymbol{\xi}}\right)=\sqrt{{\rm Tr}\left\{\left[\widetilde{\J}^{-1}\right]_{1:3G,1:3G}\right\}}.
\end{align}
It is shown in the Appendix~\ref{App} that, under the second-order approximation of the DMA analog BF matrix $\W_{\rm TX}$ in~\eqref{eq: sec_order}, the elements of the FIM in \eqref{eq: FIM}, and consequently the PEB in \eqref{eq: PEB}, can be reformulated as affine functions of either the covariance matrix $\X \triangleq \F\F^{\rm H}$ associated with the TX digital beamforming matrix or a lifted matrix formed from the real-valued reactive tuning vector $\c$.


\subsection{Problem Formulation}\label{Sec: opt_opt}
As previously mentioned, the considered bistatic sensing system possesses the SP's position uncertainty regions $\mathcal{X}_g$ $\forall g$, which implies that, for each $\mathbf{p}_g$ within $\widetilde{\boldsymbol{\xi}}$, it holds  that $\mathbf{p}_g\in\mathcal{X}_g$. Capitalizing on this available knowledge\footnote{It is noted that $\mathcal{X}_g$'s can be the result of a hypothesis, instead of the output of a dedicated tracking scheme~\cite{mendrzik2019enabling}. Intuitively, the more relevant to the actual $\mathbf{p}_g$'s those regions are, the more efficient the proposed TX DMA design will be. In the performance evaluation section that follows, we shed light onto the role of those regions' characteristics to the overall sensing performance.} for $\widetilde{\boldsymbol{\xi}}$, we propose the following mathematical problem formulation for the design objective of the TX DMA's digital and analog BF matrices:
\begin{align}
        \mathcal{P}:\nonumber&\underset{\substack{\Q_{\rm TA},\F}}{\min} \,\,\underset{\mathbf{p}_g\in\mathcal{X}_g\,\forall g}{\max}\,\,{\rm PEB}\left(\Q_{\rm TA},\F;\widetilde{\boldsymbol{\xi}}\right)\\
        &\nonumber\,\,\text{\text{s}.\text{t}.}\,\,\|\W_{\rm TX}\F\|_{\rm F}^2\leq \frac{P_{\rm tot}}{M},\,\,\forall n=1,\ldots,N_{\rm T}.
\end{align}
As shown, the objective function of this problem targets at the minimization of the worst-case PEB performance, i.e., the largest PEB value among those for the SP potential positions lying in $\mathcal{X}\triangleq\mathcal{X}_1\cup\ldots\cup\mathcal{X}_G$. To address the continuous nature of the uncertainty region $\mathcal{X}$, which extensively complicates $\mathcal{P}$'s solution, we propose to discretize it into $P$ uniformly spaced sets of points within the $(3G+2)$-dimensional grid space defined by $\widetilde{\boldsymbol{\xi}}$, which are denoted as $\{\widetilde{\boldsymbol{\xi}}_p\}_{p=1}^P$, where $\widetilde{\boldsymbol{\xi}}_p\triangleq[\p_p^{\rm T},\zeta,\Delta t]^{\rm T}\in\mathbb{R}^{(3G+2)\times 1}$. In this notation, $\p_p^{\rm T}\in\mathbb{R}^{3G\times 1}$ contains the $p$-th set of sample points\footnote{Note that within $\{\widetilde{\boldsymbol{\xi}}_p\}_{p=1}^P$ only the target positions are discretized, and the RX's rotation $\zeta$ and clock offset $\Delta t$ remain fixed. To this end, in the subsequent estimation optimization reformulation, the minimization is performed solely with respect to the FIM elements corresponding to the sample potential target positions spanning the uncertainty regions.} for the $G$ individual uncertainty regions. In addition, to handle the coupling between the DMA matrices $\Q_{\rm TA}$ and $\F$, as well as the inherent non-convexity of $\mathcal{P}$, we adopt an alternating optimization strategy: each variable block is optimized individually while keeping the other fixed, and this process iterates until convergence.

Specifically, we commence by discretizing each uncertainty region into a finite grid \(\{\widetilde{\boldsymbol{\xi}}_p\}_{p=1}^{P}\). To this end, the robust objective in \(\mathcal{P}\) reduces to a discrete worst-case criterion over \(p=1,\ldots,P\). To convexify the inverse-FIM terms \(\widetilde{\J}(\Q_{\rm TA},\F;\widetilde{\boldsymbol{\xi}}_p)^{-1}\) $\forall p$ from~\eqref{eq: PEB} using~\eqref{eq: EFIM}, we introduce the auxiliary variables \(\epsilon_{a,p}\)'s, with \(\e_a\) being an \(a\)-th canonical basis vector of the $(3G+2) \times (3G+2)$ identity matrix, to upper bound the diagonal entries of the inverse as follows:
\begin{align}
\e_a^{\rm T}\widetilde{\J}(\Q_{\rm TA},\F;\widetilde{\boldsymbol{\xi}}_p)^{-1}\e_a \le \epsilon_{a,p}\,\,\forall a,p.
\label{eq:eps_def}
\end{align}
This constraint admits an LMI representation via the Schur complement~\cite{kay1993fundamentals,keskin2022optimal}, namely:
\begin{align}
  \epsilon_{a,p}\geq 0\,\,\text{and}\,  \begin{bmatrix}
        \widetilde{\J}(\Q_{\rm TA},\F;\widetilde{\boldsymbol{\xi}}_p) & \e_a\\
        \e_a^{\rm T} & \varepsilon_{a,p}
        \end{bmatrix}\succeq 0\,\,\forall a,p.
\end{align}
Rather than minimizing \(\epsilon_{a,p}\)'s separately, which may overemphasize directions that are easier to improve, we adopt an epigraph reformulation with a single slack variable \(t\), and enforce a uniform worst-case control across all components by imposing \(\sum_{p=1}^{P}\epsilon_{a,p}\le t\) \(\forall a=1,\ldots,3G\)~\cite{kay1993fundamentals}. Next, to bypass the nonconvexity of the resulting problem due to the coupling between \(\Q_{\rm TA}\) and \(\F\) within the inverse-FIM expressions, we exploit the second-order approximation of \(\W_{\rm TX}\) in~\eqref{eq: dma_approx} to rewrite the those expressions in forms that are affine in the lifted covariance variables \(\X\) and \(\C\); this process is described in the Appendix~\ref{App}. We denote the resulting parametrizations by \(\widetilde{\J}(\X;\widetilde{\boldsymbol{\xi}}_p)\) and \(\widetilde{\J}(\C;\widetilde{\boldsymbol{\xi}}_p)\) \(\forall p\), respectively. Under semi-definite relaxation (i.e., \(\X,\C\succeq\mathbf{0}\) and temporarily ignoring rank constraints), each block subproblem becomes convex when the other block is considered fixed. Hence, instead of directly solving problem \(\mathcal{P}\), we employ an alternating  procedure that iteratively solves the following two Semi-Definite Programs (SDPs):

\vspace{-0.1cm}
\begin{align}
        \mathcal{P}_1:\nonumber&\underset{\substack{\X,\{\varepsilon_{a,p}\}_{\forall a,p},t}}{\min} \,\,t\\
        &\nonumber\text{\text{s}.\text{t}.}\,\begin{bmatrix}
        \widetilde{\J}(\X;\widetilde{\boldsymbol{\xi}}_p) & \e_a\\
        \e_a^{\rm T} & \varepsilon_{a,p}
        \end{bmatrix}\succeq 0\,\,\forall p=1,\ldots,P, 
        \\&\nonumber\,\quad \varepsilon_{a,1}+\ldots+\varepsilon_{a,P}\leq t\,\,\forall a=1,\ldots,3G,
         \\&\nonumber\,\quad  {\rm Tr}\{\W_{\rm TX}^{\rm H}\W_{\rm TX}\X\} \leq \frac{P_{\rm tot}}{M},\, \X\succeq0,
\end{align}
\begin{align}
        \mathcal{P}_2:\nonumber&\underset{\substack{\C,\{\varepsilon_{a,p}\}_{\forall a,p},t}}{\min} \,\,t\\
        &\nonumber\text{\text{s}.\text{t}.}\,\begin{bmatrix}
        \widetilde{\J}(\C;\widetilde{\boldsymbol{\xi}}_p) & \e_a\\
        \e_a^{\rm T} & \varepsilon_{a,p}
        \end{bmatrix}\succeq 0 \,\,\forall p=1,\ldots,P, 
        \\&\nonumber\,\quad \varepsilon_{a,1}+\ldots+\varepsilon_{a,P}\leq t \,\,\forall a=1,\ldots,3G,
        \\&\nonumber\,\quad  {\rm Tr}\{\V\C\}\leq\frac{P_{\rm max}}{M},\,[\C]_{N_{\rm T}+1,N_{\rm T}+1}=1,\, \C\succeq0.
\end{align}

In $\mathcal{P}_1$ and $\mathcal{P}_2$, the transmit power constraint of $\mathcal{P}$ is recast into convex trace forms with respect to the lifted variables $\X$ and $\C$, respectively.  For the analog update in $\mathcal{P}_2$, this is enabled by the fact that, under \eqref{eq: sec_order}, $\W_{\rm TX}\F$ is affine with respect to the tuning vector $\bar{\c}$. Hence, the corresponding transmit power constraint becomes a quadratic function of $\bar{\c}$, which can be equivalently written as a linear trace function of the lifted matrix $\C$, where $\V$ collects the fixed terms induced by $\W_{\rm MC}$, $\P_{\rm SA}$, and $\F$. This formulation results in convex SDPs, which can be efficiently solved using standard off-the-shelf convex optimization solvers~\cite{cvx}. 

To this end, let the optimized solutions of $\mathcal{P}_1$ and $\mathcal{P}_2$ be denoted by $\X_{\rm opt}$ and $\C_{\rm opt}$, respectively. The corresponding optimized BF matrices $\F_{\rm opt} \in \Compl^{N_{\rm RF} \times M}$ and $\c_{\rm opt} \in \Compl^{N_{\rm T} \times 1}$ can then be derived as follows. An approximate solution to $\mathcal{P}_1$ is obtained by sampling as $\F_{\rm opt} \sim \mathcal{N}(\boldsymbol{0}_{N_{\rm RF}\times M}, \X_{\rm opt})$, which satisfies\footnote{It is noted that, when $M \leq N_{\rm RF}$, a low-rank approximation for $\F_{\rm opt}$ can be derived via singular value decomposition of $\X_{\rm opt}$.} $\mathbb{E}\{\F_{\rm opt}\F_{\rm opt}^{\rm H}\} = M\X_{\rm opt}$~\cite{luo2010semidefinite}. On the other hand, the vector $\c_{\rm opt}$ is equal to $[\C_{\rm opt}]_{1:N_{\rm T},\,N_{\rm T}+1}$, since, when the SDR is tight, $\C_{\rm opt}$'s last column in our lifted optimization variable coincides with $[\c_{\rm opt}^{\rm T},1]^{\rm T}$~\cite{boyd2004convex}. However, as will be discussed in Section~\ref{Sec: Comp} later on, problems $\mathcal{P}_1$ and $\mathcal{P}_2$ pose significant computational challenges due to the large number of involved optimization variables and Linear Matrix Inequalities (LMIs) constraints required for each grid point. As a result, directly solving them may be impractical in terms of complexity. To address this issue, we next present two low complexity alternative designs: one based on a codebook-based BF strategy and the other on an approximation for $\mathcal{P}$'s CRB-based optimization objective.

\subsubsection{Codebook-Based Solution}\label{Sec: DMA_codebook_design}
It can be easily confirmed that both $\mathcal{P}_1$ and $\mathcal{P}_2$ involve large numbers of optimization variables per LMI. In particular, $N_{\rm RF}^2$ in $\mathcal{P}_1$ and $(N_{\rm T}+1)^2$ in $\mathcal{P}_2$, along with more than $3GP$ LMI constraints that must be satisfied in each problem formulation. Furthermore, due to the presence of the multiple grid points $\{\widetilde{\boldsymbol{\xi}}_p\}_{p=1}^P$, neither $\mathcal{P}_1$ nor $\mathcal{P}_2$ can be simplified using subspace reduction techniques, such as those proposed in \cite{fascista2022ris}. To overcome this challenge and provide a lower complexity alternative to the SDPs $\mathcal{P}_1$ and $\mathcal{P}_2$, we propose a codebook-driven selection strategy inspired by the methodologies in \cite{li2007range, fascista2022ris, keskin2022optimal}. In particular, building on the digital codebook introduced in~\cite[eq. (18)]{keskin2022optimal}, which was originally designed for two-Dimensional (2D) scenarios, we first extend it to the 3D case and then approximate it to construct a DMA-compatible codebook satisfying the structural constraints of this XL antenna array architecture. This codebook then constitutes the predefined dictionary of DMA beams, which enables us to cast $\mathcal{P}$ as a beam power allocation problem. This problem requires particularly the determination of the amount of power needed to allocate to each beam of the codebook across the $M$ transmission frames. Note that, since a single beam is transmitted per frame, the number of frames $M$ scales proportionally with the size of the codebook, i.e., its total number of beams. To this end, based on~\cite{li2007range} and~\cite[Proposition 1]{fascista2022ris}, we consider the 3D digital codebook $\B_{\rm dig}\triangleq\left[\B^{\rm sum},\B^{\rm diff}_{\rm \theta_{\rm a}},\B^{\rm diff}_{\rm \theta_{\rm e}}\right]\in\Compl^{N_{\rm T}\times M}$ to span the uncertainty region $\mathcal{X}$, where $\B^{\rm sum}\triangleq\left[\B^{\rm sum}_1,\ldots,\B^{\rm sum}_G\right]$, $\B^{\rm diff}_{\rm \theta_{\rm a}}\triangleq\left[\B^{\rm diff}_{{\rm \theta_{\rm a}},1},\ldots,\B^{\rm diff}_{{\rm \theta_{\rm a}},G}\right]$, and $\B^{\rm diff}_{\rm \theta_{\rm e}}\triangleq\left[\B^{\rm diff}_{{\rm \theta_{\rm e}},1},\ldots,\B^{\rm diff}_{{\rm \theta_{\rm e}},G}\right]$ constituting respectively of the following matrices:
\begin{align}
\hspace{-0.3cm}[\B_g^{\rm sum}]_{:,n_g(i,j)} \triangleq \a_{\rm TX}(\theta_{{\rm a},g,i},\theta_{{\rm e},g,j})
\end{align}
\begin{align}
[\B_{\theta_{\rm a},g}^{{\rm diff}}]_{:, n_g(i,j)} \triangleq
\frac{\partial \a_{\rm TX}(\theta_{{\rm a},g,i},\theta_{{\rm e},g,j})}{\partial\theta_{{\rm a},g,i}}
\end{align}
\begin{align}
[\B_{\theta_{\rm e},g}^{{\rm diff}}]_{:, n_g(i,j)} \triangleq
\frac{\partial \a_{\rm TX}(\theta_{{\rm a},g,i},\theta_{{\rm e},g,j})}{\partial\theta_{{\rm e},g,j}}
\end{align}
where $n_g(i,j)\triangleq(i-1)N_{{\rm \theta_{\rm e}},g}+j$ $\forall i=1,\ldots,N_{{\rm \theta_{\rm a}},g}$ and $\forall j=1,\ldots,N_{{\rm \theta_{\rm e}},g}$ with $N_{{\rm \theta_{\rm a}},g}$ and $N_{{\rm \theta_{\rm e}},g}$ being the number of beams along each angular direction, totaling $N_{{\rm beams},g}\triangleq N_{{\rm \theta_{\rm a}},g} N_{{\rm \theta_{\rm e}},g}$ beams, and $\{\theta_{{\rm a},g,i}\}_{i=1}^{N_{{\rm \theta_{\rm a}},g}}$ and $\{\theta_{{\rm e},g,j}\}_{j=1}^{N_{{\rm \theta_{\rm e}},g}}$ denote the evenly spaced AoDs for each $g$-th propagation path (i.e., contributed by the $g$-th SP). Note that $\B^{\rm sum}$ represents a standard directional beam codebook, while $\B^{\rm diff}_{\rm \theta_{\rm a}}$ and $\B^{\rm diff}_{\rm \theta_{\rm e}}$,  are derivative codebooks for each angular direction stemming from \cite[Proposition 1]{fascista2022ris}.

Realizing the fully digital codebook \(\B_{\rm dig}\) with the considered DMA architecture and its EM-compliant modeling is, in general, infeasible, as it will be next demonstrated in Section~\ref{Sec: num}. To deal with this issue, we adopt \cite{song2015codebook}'s approach and project $\B_{\rm dig}$ onto the DMA-feasible BF set, resulting in the following optimization problem:
\begin{align}
        \mathcal{P}_3&:\nonumber\underset{\substack{\Q_{\rm TA},\F}}{\min} \quad \left\|\B_{\rm dig}-(\W_{\rm MC}^{-1}-\W_{\rm MC}^{-1}\Q_{\rm TA}\W_{\rm MC}^{-1})\P_{\rm SA}\F\right\|_{\rm F}^2
\end{align}
By adopting an alternating optimization approach keeping first $\Q_{\rm TA}$ fixed, the least-squares solution $\F_{\rm opt} = (\W_{\rm TX}^{\rm H}\W_{\rm TX})^{-1}\W_{\rm TX}^{\rm H}\B_{\rm dig}$ can be obtained \cite{alexandropoulos2025extremely}. Then, for a fixed digital BF matrix $\F$, problem $\mathcal{P}_3$ can be reformulated as a least-squares problem with respect to the tunable parameters $\c$, i.e., $\c_{\rm opt}\triangleq\argmin_{\c\in\mathbb{R}^{\rm N_T}}\, \|\d-\U\c\|^2$, where $\d\triangleq\operatorname{vec}\left(\B_{\rm dig}-\S+j\W_{\rm MC}^{-1}\operatorname{diag}(\boldsymbol{\nu})\S\right)$, with $\S\triangleq\W_{\rm MC}^{-1}\P_{\rm SA}\F$, and the $i$-th column of $\U$ is given by $[\U]_{:,i}
\triangleq-j\operatorname{vec}\left(\W_{\rm MC}^{-1}\e_i\e_i^{\rm T}\S\right)$. Since $\c$ is real-valued, the least-squares solution is obtained as $\c_{\rm opt}=\left(\Re\{\U^{\rm H}\U\}\right)^{-1}
\Re\{\U^{\rm H}\d\}$.

The resulting DMA-compatible hybrid analog and digital codebook, $\B_{\rm dma}\in\Compl^{N_{\rm T}\times M}$, can be constructed as follows:
\begin{align}
    \B_{\rm dma}\triangleq&\left(\W_{\rm MC}^{-1}-\W_{\rm MC}^{-1}\jmath{\rm diag}\left(\c_{\rm opt}-\boldsymbol{\nu}\right)\W_{\rm MC}^{-1}\right)\nonumber\\
    &\times\P_{\rm SA}\F_{\rm opt}.
\end{align}
In Appendix~\ref{AppB}, it is shown that the structures of the digital codebook $\B_{\rm dig}$ and the DMA one $\B_{\rm dma}$ are both directly derived from the structure of the optimal covariance matrix in the single-point instance of problem $\mathcal{P}$.


Finally, following~\cite{keskin2022optimal}, once the DMA-compatible beam codebook \(\B_{\rm dma}\) is constructed with \(M \triangleq 3\sum_{g=1}^{G} N_{{\rm beams},g}\) codewords (recall that \(M\) also equals the number of transmission frames required to span the uncertainty region \(\mathcal{X}\)), the BF design problem \(\mathcal{P}\), under the relaxations used in \(\mathcal{P}_1\) and \(\mathcal{P}_2\), can be recast as a beam power allocation problem over \(\boldsymbol{\rho}\triangleq[\rho_1,\ldots,\rho_M]^{\rm T}\in\mathbb{R}^{M}\), where \(\rho_m\) denotes the power assigned to codeword \([\B_{\rm dma}]_{:,m}\) in each \(m\)-th frame, i.e.:

\begin{align}
        \mathcal{P}_5:\nonumber&\underset{\substack{\boldsymbol{\rho},\{\varepsilon_{a,p}\}_{\forall a,p},t}}{\min} \,\,t\\
        &\nonumber\text{\text{s}.\text{t}.}\,\begin{bmatrix}
        \widetilde{\J}(\Z;\widetilde{\boldsymbol{\xi}}_p) & \e_a\\
        \e_a^{\rm T} & \varepsilon_{a,p}
        \end{bmatrix}\succeq 0\,\,\forall p=1,\ldots,P, 
        \\&\nonumber\,\quad \varepsilon_{a,1}+\ldots+\varepsilon_{a,P}\leq t \,\,\forall a=1,\ldots,3G,
        \\&\nonumber\,\quad  {\rm Tr}\{\Z\} \!\leq\! \frac{P_{\rm tot}}{M},\,\Z\!=\!\B_{\rm dma}{\rm diag}(\boldsymbol{\rho})\B^{\rm H}_{\rm dma},\,\boldsymbol{\rho}\!\geq\!\boldsymbol{0}_{\rm M_{\rm R}\times 1},
\end{align}
where $\widetilde{\J}(\Z;\widetilde{\boldsymbol{\xi}}_p)$ $\forall p$ is the FIM in \eqref{eq: EFIM} reformulated as in \eqref{eq: FIM_lin_X} with respect to $\Z\triangleq\W_{\rm TX}\X\W_{\rm TX}^{\rm H}$. This leads to a convex formulation that can be efficiently solved using standard off-the-shelf solvers, yielding the optimal solution $\boldsymbol{\rho}_{\rm opt}$. From this solution, the overall codebook-based BF strategy solving $\mathcal{P}$ can be finally designed as $\B_{\rm dma}\text{diag}(\sqrt{\boldsymbol{\rho}_{\rm opt}})$.

\subsubsection{An Upper Bound Solution}\label{Sec: LB} 
Starting from the FIM expression in \eqref{eq: EFIM} and considering the inequality ${\rm Tr}\{\widetilde{\J}^{-1}\}\geq{\rm Tr}\{\widetilde{\J}\}^{-1}$, which follows from the harmonic-geometric mean inequality~\cite{gavras2025near}, we propose to neglect the interdependence captured in FIM's off-diagonal elements, resulting in a lower bound approximation for it. To this end, instead of directly minimizing the CRB at each grid point, as in $\mathcal{P}$, we focus on indirectly minimizing the worst-case PEB via the latter bound. This upper bound approach eliminates all LMI constraints appearing in~$\mathcal{P}_5$, thus, significantly simplifying that design optimization formulation, which is now expressed as follows:
\begin{align}
        \mathcal{P}_6:\,\nonumber&\underset{\substack{\boldsymbol{\rho}}}{\max} \,\,\sum_{a=1}^{3G}\sum_{p=1}^P\e_a^{\rm T}\widetilde{\J}(\Z;\widetilde{\boldsymbol{\xi}}_p)\e_a\\
        &\nonumber\text{\text{s}.\text{t}.}\,{\rm Tr}\{\Z\} \!\leq\! \frac{P_{\rm tot}}{M},\, \Z\!=\!\B_{\rm dma}{\rm diag}(\boldsymbol{\rho})\B_{\rm dma}^{\rm H},\,\boldsymbol{\rho}\!\geq\!\boldsymbol{0}_{\rm M_{\rm R}\times1}.
\end{align}
This leads to a convex formulation that can be efficiently solved using the same approach as in $\mathcal{P}_5$, yielding  with the optimized solution recovered in an analogous manner.

\subsection{Complexity Analysis}\label{Sec: Comp}
In SDPs, the main computational bottleneck typically stems from the large number of LMIs whose complexity scales with both the number of optimization variables and the dimension of each LMI constraint. For $\mathcal{P}$'s reformulations, a significant number of LMIs is involved, with complexity growing with both the numbers of BS antennas $N_{\rm T}$ and SPs $G$. Using interior-point methods, the worst-case complexity for solving SDPs is $\mathcal{O}(z^2\sum_{i=1}^C b_i^2 + z\sum_{i=1}^C b_i^3)$~\cite{nemirovski2004interior,boyd2004convex}, where $z$ is the number of optimization variables, $C$ is the number of LMI constraints, and $b_i$ is the dimension of each $i$-th LMI constraint.
For the direct CRB minimization approach, i.e., iteratively solving $\mathcal{P}_1$ and $\mathcal{P}_2$, it holds that: for $\mathcal{P}_1$, we have $z = N_{\rm RF}^2 + 3GP + 1$ and $C = 3GP + 1$ with $b_i = 3G + 3$ for $i < C$ and $b_C = N_{\rm RF}$. Assuming $P \ll N_{\rm T}$, a common scenario in practical deployments, the resulting complexity is approximately $\mathcal{O}(N_{\rm RF}^2N_{\rm T}^4)$. Similarly, for $\mathcal{P}_2$, we have $z = N_{\rm T}^2$ and $C = 3GP + 1$ with $b_i = 3G + 3$ for $i < C$ and $b_C = N_{\rm T}$, yielding a worst-case complexity of $\mathcal{O}(N_{\rm T}^6)$. Thus, the overall complexity for the iterative CRB minimization scheme is $\mathcal{O}\left(I_{\rm max}(N_{\rm RF}^2N_{\rm T}^4+N_{\rm T}^6)\right)$, where $I_{\rm max}$ denotes the maximum number of iterations required for convergence. Additionally, the generation of the codebook $\B_{\rm dma}$ via $\mathcal{P}_4$, using the conjugate gradient method, incurs an approximate complexity of $\mathcal{O}(N_{\rm T}^{1.5} N_{\max})$ \cite{sato2022riemannian}, where $N_{\max}$ denotes the maximum number of iterations until convergence; we have assumed that the gradient evaluation cost is negligible relative to the problem’s dimensionality. Finally, following a similar reasoning to that for $\mathcal{P}_1$ and $\mathcal{P}_2$, solving $\mathcal{P}_5$, under the assumption that $G \ll M$, 
yields a computational complexity of $\mathcal{O}(M^5)$, whereas for the lower-bound approximation approach in $\mathcal{P}_6$, the computational complexity is deduced to $\mathcal{O}(M^3)$.


\section{Target Estimation with DMA Beamforming}
\label{sec:ml}
In this section, we present a multi-target parameter estimation approach for the optimized DMA-based bistatic sensing system described in Section~\ref{Sec: Opt}. 
We first derive the log-likelihood function in terms of the parameters of interest and the nuisance ones, and then devise a sequential estimator that alternates among parameter updates to recover all unknowns.

\subsection{Observation Signal Model}
\label{sec:ml_compact_model}
%
%
Starting from \eqref{eq: rec_signal}, we formulate the $M_{\rm R}\times T$ complex-valued matrix $\mathbf Y_{m,k} \triangleq [\mathbf y_{m,k,1},\ldots,\mathbf y_{m,k,T}]$ with the \(T\) received pilots at the \(k\)-th SC via the \(m\)-th TX precoded frame (i.e., $s_{m,k,t}$ $\forall t$). After removing the known pilot symbols, the following $M_{\rm R}$-element observation signal is obtained $\forall m,k$:
\begin{equation}
    \widetilde{\mathbf y}_{m,k}
    \triangleq
    \frac{1}{\sqrt{T}}\mathbf Y_{m,k}\left(\left[\S_m\right]_{k,:}\right)^*.
    \label{eq:pilot_comp}
\end{equation}
Consequently, we define the global observation vector $ \bar{\mathbf y}
    \triangleq
    \big[\widetilde{\mathbf y}_{1}^{\rm T},\ldots,\widetilde{\mathbf y}_{M}^{\rm T}\big]^{\rm T}
    \in\mathbb C^{M K M_{\rm R}\times 1}$, where each vector $\widetilde{\mathbf y}_{m}
    \triangleq
    \big[\widetilde{\mathbf y}_{m,1}^{\rm T},\ldots,\widetilde{\mathbf y}_{m,K}^{\rm T}\big]^{\rm T}
    \in\mathbb C^{K M_{\rm R}\times 1}$ includes all observed signals at the \(m\)-th frame from all $K$ SCs.

We also define the following $KM_{\rm R}$-element complex-valued vector for each $(m,g)$-th channel path:
\begin{align}
    \mathbf{g}_{m,g}(\mathbf p_g,\zeta,\Delta t)
    \triangleq
    \begin{bmatrix}
        e^{-j2\pi f_1 \tau_g(\mathbf p_g,\Delta t)}\widetilde{\a}_m(\p_g,\zeta)\\
        \vdots\\
        e^{-j2\pi f_K \tau_g(\mathbf p_g,\Delta t)}\widetilde{\a}_m(\p_g,\zeta)
    \end{bmatrix},
    \label{eq:gm_g_def}
\end{align}
where, using notation $\boldsymbol\psi_g(\mathbf p_g)$ for the geometry-dependent AoA parametrization (i.e., for both azimuth and elevation) and $\boldsymbol\theta_g(\mathbf p_g,\zeta)$ for the corresponding (geometry- and yaw-dependent) AoD parametrization, $\widetilde{\a}_m(\p_g,\zeta)$ is expressed as:
\begin{align}\nonumber
    \widetilde{\a}_m(\p_g,\zeta)\triangleq\mathbf W_{\rm RX}^{\rm H}\mathbf a_{\rm RX}\big(\boldsymbol\psi_g(\mathbf p_g)\big)\mathbf a_{\rm TX}^{\rm H}\big(\boldsymbol\theta_g(\mathbf p_g,\zeta)\big)\mathbf W_{\rm TX}\mathbf f_m.
\end{align}
All $G$ latter vectors can be then used to formulate $\mathbf G_m(\widetilde{\boldsymbol{\xi}})
    \triangleq
    \big[\mathbf g_{m,1}(\p_1,\zeta,\Delta t),\ldots,\mathbf g_{m,G}(\p_1,\zeta,\Delta t)\big]\in\mathbb{C}^{KM_{\rm R}\times G}$, concerning each \(m\)-th transmission frame, and in sequel, by stacking all $M$ frames, we formulate the global steering matrix $\mathbf G(\widetilde{\boldsymbol{\xi}})
    \triangleq
    \big[\mathbf G_1^{\rm T}(\widetilde{\boldsymbol{\xi}}),\ldots,\mathbf G_M^{\rm T}(\widetilde{\boldsymbol{\xi}})\big]^{\rm T}
    \in\mathbb C^{M K M_{\rm R}\times G}$.

Finally, by grouping all path gains into the $G$-element vector definition $\mathbf a\triangleq[a_1,\ldots,a_G]^{\rm T}$, the pilot-free observation signal at the RX's output RF chains can be compactly expressed as:
\begin{equation}\label{eq:glogal_observation}
\bar{\mathbf{y}}=\G(\widetilde{\boldsymbol{\xi}})\a+ \bar{\n},
\end{equation}
where $\bar{\n} \triangleq
    \big[\widetilde{\mathbf n}_{1}^{\rm T},\ldots,\widetilde{\mathbf n}_{M}^{\rm T}\big]^{\rm T}
    \in\mathbb C^{M K M_{\rm R}\times 1}$, with $\widetilde{\mathbf n}_{m}
    \triangleq
    \big[\widetilde{\mathbf n}_{m,1}^{\rm T},\ldots,\widetilde{\mathbf n}_{m,K}^{\rm T}\big]^{\rm T}
    \in\mathbb C^{K M_{\rm R}\times 1}$, is the noise vector from \eqref{eq: rec_signal}.

\subsection{Concentrated Maximum-Likelihood Estimation}
\label{sec:ml_profiled}
Conditioned on $\widetilde{\boldsymbol{\xi}}$, including
the unknown channel and location parameters (i.e., $\widetilde{\boldsymbol{\xi}}\triangleq[\p^{\rm T},\zeta,\Delta t]^{\rm T}$), the observation vector $\bar{\y}$ in~\eqref{eq:glogal_observation} is a linear Gaussian function of the nuisance vector $\a$ \cite{millar2011maximum}. Consequently, the conditional log-likelihood is: 
\begin{equation}
    \log p\!\left(\bar{\mathbf y}\mid \mathbf p,\zeta,\Delta t\right)
    =
    -\frac{1}{\sigma^2}
    \left\|
        \bar{\y}
        -
        \G(\mathbf p,\zeta,\Delta t)\a
    \right\|^2.
    \label{eq:ll_joint}
\end{equation}
For fixed $(\mathbf p,\zeta,\Delta t)$, the ML estimate of $\mathbf a$ coincides with the Least-Squares (LS) solution, which is expressed as follows:
\begin{align}
    \widehat{\a}
    &=
    \argmin_{\a}
    \left\|
        \bar{\mathbf y}
        -
        \G(\mathbf p,\zeta,\Delta t)\a
    \right\|^2=
    \big(\mathbf G^{\rm H}\mathbf G\big)^{-1}\mathbf G^{\rm H}\bar{\y},
    \label{eq:a_hat_ls}
\end{align}
where, for brevity\footnote{Note that the concentrated log-likelihood is identical when computing an LS estimate of the full gain vector \(\a\) and of \(\beta_g\)'s. This happens because the corresponding projection matrices span the same column space when the RX does not exploit any additional structure on the latter.}, $\mathbf G$ implies $\mathbf G(\mathbf p,\zeta,\Delta t)$ . Substituting \eqref{eq:a_hat_ls} into \eqref{eq:ll_joint}, yields the concentrated log-likelihood function:
\begin{equation}
    \mathcal{L}(\mathbf p,\zeta,\Delta t)
    \triangleq
    \log p\!\left(\bar{\mathbf y}\mid \mathbf p,\zeta,\Delta t\right)\Big|_{\a=\widehat{\a}}
    \propto
    \left\|
        \P_{\G}\,\bar{\y}
    \right\|^2,
    \label{eq:profiled_L}
\end{equation}
where $\P_{\G}\triangleq\G\big(\G^{\rm H}\G\big)^{-1}\G^{\rm H}$ denotes the orthogonal projector onto the column space of $\G$. Finally, the ML estimation of the unknown parameters included in $\widetilde{\boldsymbol\xi}$ is obtained as follows:
\begin{equation}
    \big(\mathbf p_{\rm ML},\zeta_{\rm ML},\Delta t_{\rm ML}\big)
    \triangleq
    \argmax_{\mathbf p,\zeta,\Delta t}
    \mathcal{L}(\mathbf p,\zeta,\Delta t).
    \label{eq:theta_ml}
\end{equation}
A brute-force maximization of \eqref{eq:theta_ml} is typically intractable due to the high dimensionality and nonconvexity of function $\mathcal{L}(\cdot)$. In the sequel, we present a sequential solution for this problem that remains consistent with the profiled ML principle in \eqref{eq:profiled_L}.

\subsection{Sequential Parameter Estimation}
\label{sec:ml_param_est}
To solve~\eqref{eq:theta_ml} including the concentrated log-likelihood objective, we follow a three-step sequential estimation approach, solving for each optimization variable individually while keeping the remaining two variables fixed, as follows.

\subsubsection{Target Positions}
\label{sec:sp_sequential}
The target positions are estimated sequentially via the approach: first, \textit{i}) detection of the most dominant path; then, \textit{ii}) assignment of this path to the most likely uncertainty region, followed by \textit{iii}) refinement of its position within that region; and finally, \textit{iv}) subtraction of that path's contribution from the received signal. This procedure is repeated for each of the $G$ targets.

Suppose the availability of the estimates $\widehat{\zeta}$ and $\widehat{\Delta t}$ for the parameters $\zeta$ and $\Delta t$, respectively, and the initialization $\r^{(1)} \triangleq \bar{\y}$. Let also $\mathbf{u}\in\mathbb{R}^{3\times 1}$ indicate a candidate target position, for which we define the corresponding global steering vector $\widetilde{\g}\big(\mathbf u,\widehat{\zeta},\widehat{\Delta t}\big)
    \triangleq
    \big[
      \mathbf g_{1}^{\rm T}\big(\mathbf u,\widehat{\zeta},\widehat{\Delta t}\big),\ldots,
      \mathbf g_{M}^{\rm T}\big(\mathbf u,\widehat{\zeta},\widehat{\Delta t}\big)
    \big]^{\rm T}$, 
where each $\mathbf g_{m}\big(\mathbf u,\widehat{\zeta},\widehat{\Delta t}\big)$ is obtained as in~\eqref{eq:gm_g_def} by substituting $(\p_g,\zeta,\Delta t)$ with $\big(\mathbf u,\widehat{\zeta},\widehat{\Delta t}\big)$. For each $g$-th target ($g=1,\ldots,G$) accompanied with the uncertainty region $\mathcal{X}_g$, we calculate the normalized matched filtering score $\forall\u\in\mathcal{X}_g$:
\begin{equation}
    J_g(\u)
    \triangleq
    \frac{
      \big|
        \widetilde{\g}^{\rm H}\big(\mathbf u,\widehat{\zeta},\widehat{\Delta t}\big)\,\mathbf r^{(g)}
      \big|^2
    }{
      \big\|
        \widetilde{\g}\big(\mathbf u,\widehat{\zeta},\widehat{\Delta t}\big)
      \big\|^2
    }.
    \label{eq:mf_metric_region}
\end{equation}
Then, the following problem providing a coarse target position estimation $\widehat{\mathbf p}_g$, yielding the strongest propagation component, and the associated uncertainty region with this target (via the estimated region index $\widehat{\ell}_g$) is solved as (with $\mathcal{G}\subseteq\{1,\ldots,G\}$):
\begin{equation}
    (\widehat{\ell}_g,\widehat{\mathbf p}_g)
    \triangleq
    \argmax_{\ell\in\mathcal{G}}
    \ \max_{\u\in\mathcal{X}_{\ell}}
    J_g(\u).
    \label{eq:select_region_position}
\end{equation}
The resulting coarse estimate $\widehat{\mathbf p}_g$ is then refined by maximizing \eqref{eq:mf_metric_region} over $\mathcal{X}_{\widehat{\ell}_g}$ using a finer spatial resolution. The resolution of this fine search can be actually selected based on the CRB associated with this coarse position estimate. Given $\widehat{\ell}_g$ and the refined $\widehat{\mathbf p}_g$, the corresponding complex attenuation coefficient resulting from the signal reflection from the $g$-th target can be estimated via the scalar LS projection, as follows:
\begin{equation}
    \widehat{a}_g
    =
    \frac{
      \mathbf g^{\rm H}(\widehat{\mathbf p}_g,\zeta,\Delta t)\,\mathbf r^{(g)}
    }{
      \big\|
        \mathbf g(\widehat{\mathbf p}_g,\zeta,\Delta t)
      \big\|^2
    }.
    \label{eq:aq_ls}
\end{equation}

Finally, the residual for the estimation of the parameters of the $(g+1)$-th target is updated by subtracting the estimated contribution of the $g$-th target from the observation signal, yielding $\r^{(g+1)}=\r^{(g)}-\widetilde{\g}\big(\widehat{\mathbf p}_g,\widehat{\zeta},\widehat{\Delta t}\big)\,\widehat{a}_g$. In addition, the selected uncertainty region index $\widehat{\ell}_g$ is removed from the set $\mathcal{G}$ (i.e., $\mathcal{G}\leftarrow\mathcal{G}\setminus\{\widehat{\ell}_g\}$), implying that further association for that $g$-th target is unnecessary.  

\subsubsection{Clock Offset}
\label{sec:ml_dt}
Given the estimates $\widehat{\mathbf p}$ and $\widehat{\zeta}$, the clock offset can be estimated by an One-Dimensional (1D) search over a plausible interval $\mathcal T$, as follows:
\begin{equation}
    \widehat{\Delta t}
    =
    \argmax_{\Delta t\in\mathcal T}
    \mathcal{L}\big(\widehat{\p},\widehat{\zeta},\Delta t\big).
    \label{eq:dt_update}
\end{equation}
Recall that the clock-bias variance \(\sigma_{\rm clk}^2\) is assumed known, hence, a simple initialization point can be obtained by sampling from its prior or, alternatively, via coarse delay estimation based on the power delay profile, as in~\cite{RastorguevaFoi2024mmWaveSLAM}. A compact feasible interval can be naturally set as
\(\mathcal{T}\triangleq[-\kappa\sigma_{\rm clk},\,\kappa\sigma_{\rm clk}]\),
where \(\kappa>0\) represents a user-defined confidence factor.
 
\subsubsection{RX Orientation Angle}
\label{sec:ml_zeta}
Given the estimates $\widehat{\p}$ and $\widehat{\Delta t}$, the yaw can be estimated similar to~\eqref{eq:dt_update} as:
\begin{equation}
    \widehat{\zeta}
    =
    \argmax_{\zeta\in[-0.5\pi,0.5\pi]}
    \mathcal{L}\big(\widehat{\p},\zeta,\widehat{\Delta t}\big).
    \label{eq:zeta_update}
\end{equation}
Note that \(\Delta t\) induces a common linear phase rotation across SCs, whereas $\zeta$ affects the steering vectors through a nonlinear coordinate rotation and AoD mapping. Hence, $\zeta$ can be conveniently estimated via an 1D search with optional Newton/Gauss-Newton refinement, as in \cite{burke1995gaussnewton}. A natural initialization for \(\zeta\) is the center of its feasible interval.

\subsubsection{Overall Approach and Complexity}
\label{sec:ml_alternating}
The proposed approach for the sequential estimation of $\p$, $\zeta$, and $\Delta t$ maximizing the concentrated log-likelihood objective in~\eqref{eq:theta_ml} is summarized in Algorithm~\ref{Algo}. Steps $2$-$7$ compute coarse estimations for all $G$ target positions, whereas Steps $9$-$13$, with $L_{\rm\max}$ denoting a user-defined number for the maximum iterations, perform iterative refinement of all estimates via \eqref{eq:mf_metric_region}-\eqref{eq:zeta_update} within coarse-to-fine grids over the respective uncertainty regions. 
The complexity of Algorithm~\ref{Algo} is dominated by the repeated matched filtering evaluations during the coarse-to-fine initialization and alternating refinement. Let $\bar{N}\triangleq MKM_R$ be the length of the global observation vector $\bar{\y}$ and $S$ the total number of evaluations of \eqref{eq:theta_ml}'s objective (across grid points, 1D searches, and refinements). The overall runtime is bounded by $\mathcal{O}\!\left(S(\bar{N}G^2+G^3)\right)$. In the typical regime where $G$ is small and fixed, this reduces to an essentially linear scaling with the observation data dimension, i.e., $\mathcal{O}(S\bar{N})$.


\begin{algorithm}[t]
\caption{Proposed Sequential Bistatic Sensing}
\label{alg:full_compact_refs}
\begin{algorithmic}[1]

\REQUIRE $\{\mathcal{X}_g\}_{g=1}^G$, $\mathcal T$, $\mathcal{G}$, $\bar{\y}$, $G$, and $L_{\rm\max}$.
\STATE Initialize $\r^{(1)}=\bar{\y}$, and set $\Delta t^{(0)}$ and $\zeta^{(0)}$ according to Sections~\ref{sec:ml_dt} and~\ref{sec:ml_zeta}, respectively.
\FOR{$g=1,2,\ldots,G$}
    \STATE Obtain $(\widehat{\ell}_g,\widehat{\mathbf p}_g)$ solving~\eqref{eq:select_region_position}.
    \STATE Refine $\widehat{\mathbf p}_g$ within $\mathcal X_{\widehat{\ell}_g}$ via maximizing \eqref{eq:mf_metric_region}.
    \STATE Compute $\widehat a_g$ via~\eqref{eq:aq_ls}.
    \STATE Compute residual $\r^{(g+1)}\!=\!\mathbf r^{(g)}-\mathbf g(\widehat{\mathbf p}_g,\zeta^{(0)},\Delta t^{(0)})\widehat{a}_g$.
    \STATE Update index set as $\mathcal{G}\leftarrow\mathcal{G}\setminus\{\widehat{\ell}_g\}$.
\ENDFOR
\FOR{$i=0,1,\ldots,L_{\rm\max}-1$}
    \STATE Update $\Delta t^{(i+1)}$ via~\eqref{eq:dt_update} using \(\zeta^{(i)}\) and $\p_g^{(i)}$ $\forall g$.
    \STATE Update $\zeta^{(i+1)}$ via~\eqref{eq:zeta_update} using \(\Delta t^{(i+1)}\) and $\p_g^{(i)}$ $\forall g$.
    \STATE Update \(\p_g^{(i+1)}\) \(\forall g\) via~\eqref{eq:mf_metric_region} using \(\Delta t^{(i+1)}\), \(\zeta^{(i+1)}\), $\widehat{\ell}_g$.
\ENDFOR
\STATE \textbf{Output:} $\widehat{\p}_g=\p_g^{(L_{\rm\max})}$ $\forall g$, $\widehat{\zeta}=\zeta^{(L_{\rm\max})}$, $\widehat{\Delta t}=\Delta t^{(L_{\rm\max})}$.

\end{algorithmic}
\label{Algo}
\end{algorithm}

\section{Numerical Results and Discussion}\label{Sec: num}
In this section, we present simulation results assessing the performance of the proposed DMA-enabled, mutual-coupling-aware framework for robust bistatic sensing. Specifically, we investigate how the SPs' sensing accuracy is affected by mutual coupling effects, spatial uncertainties at the SPs, and clock bias inconsistencies at the RX side.

\subsection{Simulation Parameters and Benchmarks}
To evaluate the performance of the proposed robust bistatic sensing framework, simulations have been carried out using the parameters included in Table I, unless otherwise stated, considering an environment encapsulating $G=2$ SPs. 
For the $\W_{\rm RX}$ design, we have followed an approach inspired our 3D digital codebook design in Section~\ref{Sec: DMA_codebook_design}: the first $M_{\rm R}$ left singular vectors of $\A_{\rm dig}$, scaled to have unit amplitude were selected. In this case, $\A_{\rm dig}$ represents the receive digital codebook which is structured analogously to $\B_{\rm dig}$, but it is generated using the RX steering vector $\a_{\rm RX}(\cdot)$ and the corresponding AoAs at the RX. Additionally, we have explored two scenarios in which the position uncertainty of the SPs was set to $u_g=5$~m $\forall g$ (Scenario 1) and $u_g=0.5$~m $\forall g$ (Scenario 2). Each uncertainty region $\mathcal{X}_g$ was assumed to span the azimuth and elevation angles in $[\theta^{\min}_{{\rm a}, g}, \theta^{\max}_{{\rm a},g}]$ and $[\theta^{\min}_{{\rm e},g}, \theta^{\max}_{{\rm e},g}]$, respectively. The azimuth and elevation beamwidths of the DMA-based TX has been approximated as $\Delta\theta_{\rm a}\approx\frac{\lambda}{N_{\rm E}d_x}$ and $\Delta\theta_{\rm e}\approx\frac{\lambda}{N_{\rm RF}d_y}$ \cite{balanis2016antenna}, respectively. Hence, the number of beams required to span each $g$-th uncertainty region (modeled as a sphere) along each angular dimension was set as:
\begin{align}
N_{{\rm\theta_{\rm a}},g}=\left \lceil{\frac{\theta^{\max}_{{\rm a},g}-\theta^{\min}_{{\rm a},g}}{\Delta\theta_{\rm a}}}\right \rceil,\, N_{{\rm\theta_{\rm e}},g}=\left \lceil{\frac{\theta^{\max}_{{\rm e},g}-\theta^{\min}_{{\rm e},g}}{\Delta\theta_{\rm e}}}\right \rceil.
\end{align}
Based on the above, we have obtained approximately $P = 8$ points and $M=48$ frames for Scenario 1, and approximately $P = 2$ points and $M=12$ frames for Scenario 2, to span the entirety of the uncertainty regions. To facilitate a direct connection between the clock bias uncertainty $\sigma_{\rm clk}$ and PEB, both quantities are expressed in meters. All numerical results were obtained by averaging over $500$ Monte Carlo simulations.

We have conducted a comparative evaluation of the proposed DMA-compatible BF designs for bistatic sensing against the following benchmark approaches: \textit{i}) the BF design in~\cite{gavras2025near} based on the CRB Lower Bound (LB); \textit{ii}) the BF strategy in Section~\ref{Sec: DMA_codebook_design} excluding mutual coupling effects via omitting $\W_{\rm MC}$'s contribution; and \textit{iii}) fully digital and analog BF designs in \cite{keskin2022optimal}. All methods have been assessed under identical conditions, including the number of TX RF chains, antenna elements, and inter-element spacing. Notably, our proposed framework is the only one that explicitly accounts for mutual coupling effects in the BF design analysis.

\begin{table}[!t]\label{Table 1}
\centering
\caption{Simulation Parameters.}
\begin{tabular}{|c|c|c|c|}
\hline
Parameter & Value & Parameter & Value \\
\hline\hline
$f_c$ & 24 GHz & $\p_{\rm RX}$ & (35, 6, 5) m \\
\hline
$K$ & 512 & $\p_1$ & (5 15 5) m \\
\hline
$\Delta f$ & 120 kHz & $\p_2$ & (15, 5, 5) m \\
\hline
$T$ & 1 & $\zeta$  & $\pi/4$ \\
\hline
$N_{\rm RF}$ & 4 & $a$ & $0.73\lambda$\\
\hline
$N_{\rm E}$ & 32 & $b$ & $0.17\lambda$ \\
\hline
$d_{\rm RF}, d_{\rm E}$  & $\frac{\lambda}{2}, \frac{\lambda}{5}$ & $S_{\mu}$  & $110$ mm\\
\hline
$M_{\rm R}$  & 16 & $\sigma^2$ & $-80$ (dBm) \\
\hline
$P_{\rm tot}/M$ & 20 dBm & $b_{\rm bor}$ & 0.57 \\
\hline
\end{tabular}
\end{table}

\subsection{Beamforming Performance}

\begin{figure}[!t]
	\begin{center}
	\includegraphics[width=\columnwidth]{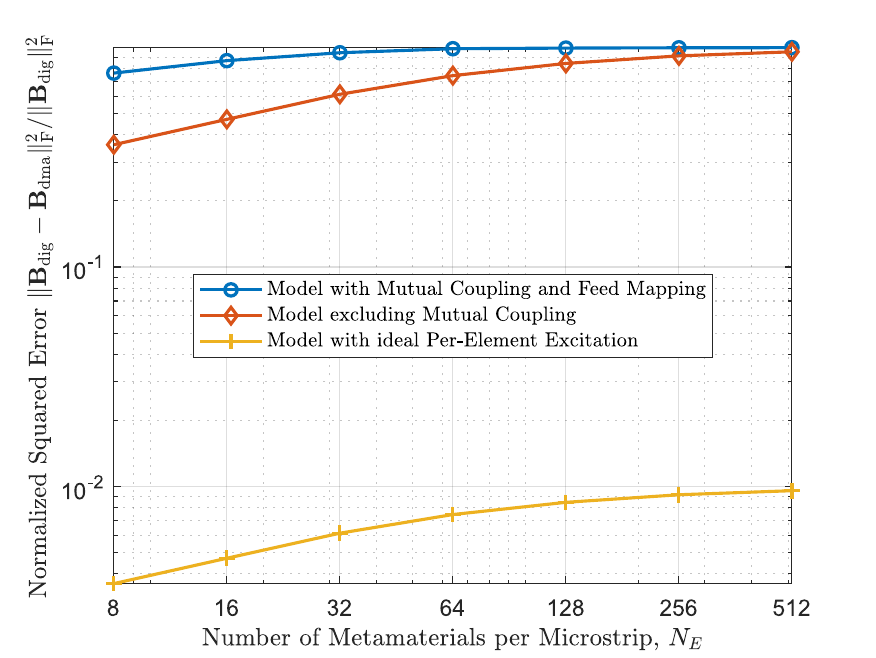}
	\caption{\small{Normalized squared norm difference between the optimized digital codebook $\B_{\rm dig}$ and the DMA codebook $\B_{\rm dma}$ as a function of the number of metamaterial elements $N_{\rm E}$ per TX RF chain, considering three scenarios: mutual coupling effects (via $\W_{\rm MC}$) and in-waveguide propagation (via $\P_{\rm SA}$) included, and exclusions of either of these two effects.}} 
    \vspace{-0.4cm}
	\label{fig: Codebook_cost}
	\end{center}
\end{figure}

To assess the performance of the proposed EM-compliant DMA-based TX BF design in Section~\ref{Sec: DMA_codebook_design} for bistatic sensing, we first quantify how accurately the fully digital codebook $\B_{\rm dig}$ can be approximated by the designed DMA-feasible (hybrid analog/digital) codebook $\B_{\rm dma}$. Figure~\ref{fig: Codebook_cost} depicts the mismatch  $\|\B_{\rm dig}-\B_{\rm dma}\|_{\rm F}^2/\|\B_{\rm dig}\|_{\rm F}^2$ versus the number of metamaterial elements $N_{\rm E}$ per guided channel. Three variants of the TX DMA response were considered in the design: \textit{i}) the full EM-compliant model in~\eqref{eq: BF}, i.e., $\W_{\rm TX}=(\Q_{\rm TA}+\W_{\rm MC})^{-1}\P_{\rm SA}$, which jointly captures propagation leakage along the feeding structure via $\P_{\rm SA}$ and mutual coupling via $\W_{\rm MC}$; \textit{ii}) a special case without mutual coupling for which $\W_{\rm TX}=\Q_{\rm TA}^{-1}\P_{\rm SA}$; and \textit{iii}) an idealized ablation in which the propagation mapping $\P_{\rm SA}$ is disabled (equivalently, ideal per-element excitation), i.e., $\W_{\rm TX}=(\Q_{\rm TA}+\W_{\rm MC})^{-1}$ (note that this case requires an adjustment of $\F$'s dimensions). As observed, none of the considered DMA models can reproduce the unconstrained fully digital beams, while the mismatch increases with $N_{\rm E}$, reflecting the stronger structural constraints induced by the guided feed and progressive leakage along the channel encoded via $\P_{\rm SA}$. It is also shown that ignoring mutual coupling improves the approximation, yet in-waveguide propagation remains a dominant limiting factor. This is further highlighted when the in-waveguide propagation is removed in the idealized ablation, the proposed design can closely approximate the fully digital beams, thereby isolating the impact of this propagation from that of mutual coupling. Similar trends to Fig.~\ref{fig: Codebook_cost} are also observed for increasing numbers of TX RF chains, $N_{\rm RF}$, and frames, $M$.

\begin{figure*}[!t]
    \centering
    \includegraphics[width=0.9\textwidth]{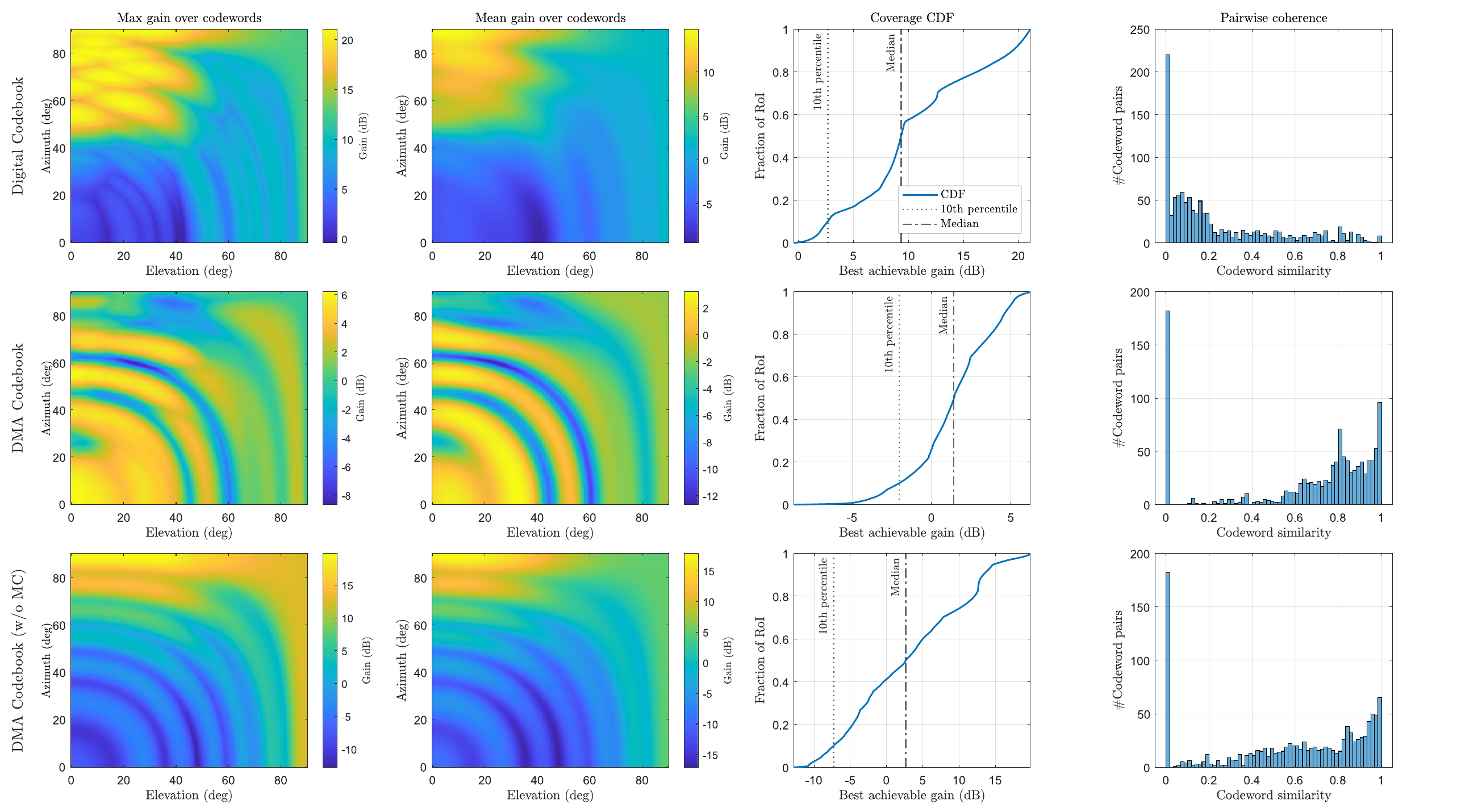}
    \caption{\small{Beamforming-gain statistics and codebook coherence for the fully-digital codebook $\B_{\rm dig}$, the proposed EM-compliant DMA codebook $\B_{\rm dma}$, and the DMA codebook obtained by neglecting mutual coupling (i.e., omitting $\W_{\rm MC}$), for Scenario~1 over the azimuth--elevation grid: maximum (best-beam) gain across codewords (\textit{first column}), mean gain across codewords (\textit{second column}), empirical CDF of the maximum gain over the Region of Interest (RoI) (\textit{third column}), and histogram of pairwise codeword coherence (\textit{fourth column}).}} 
        \vspace{-0.4cm}
  \label{fig: beams_all}
\end{figure*}

Figures~\ref{fig: beams_all} report beamforming-gain diagnostics for the fully digital codebook $\B_{\rm dig}$, the proposed EM-compliant DMA codebook $\B_{\rm DMA}$, and a DMA codebook designed by neglecting mutual coupling (i.e., excluding $\W_{\rm MC}$), for Scenario~1 of SPs' position uncertainty. In each row the four panels provide codebook-only views over the scanned azimuth--elevation field-of-view: the \textit{first} column showcases the maximum (e.g. best-beam) array gain across codewords at each look direction, revealing coverage and potential holes; the \textit{second} column reports the mean gain across codewords, indicating the average energy distribution and any angular bias; the \textit{third} column presents the empirical Cumulative Distribution Function (CDF) of the maximum gain over the Region of Interest (RoI), summarizing overall coverage; and the \textit{fourth} column depicts the histogram of pairwise codeword coherence (i.e., the normalized inner product between beam pairs), quantifying beam diversity versus redundancy. The digital benchmark $\B_{\rm dig}$ exhibits smoothly steerable responses with limited coverage holes, reflected by a right-shifted CDF and predominantly low coherence values. In contrast, the EM-compliant $\B_{\rm DMA}$ shows reduced peak gains and a more structured angular footprint due to the constrained excitation imposed by the guided-wave feed/leakage profile and mutual-coupling distortions, which manifests as a left-shifted CDF and increased high-coherence mass, indicating reduced beam diversity and, consequently, degraded sensing performance. Finally, by isolating mutual-coupling effects through omitting $\W_{\rm MC}$, we observe a partial recovery of digital-like behavior-higher peak/mean gains, a right-shifted CDF relative to the digital case, and fewer highly coherent codeword pairs-thereby underscoring the importance of EM-compliant modeling in DMA codebook design.

\subsection{Sensing and Localization Performance}
\begin{figure}[!t]
  \centering
  \includegraphics[width=0.5\textwidth]{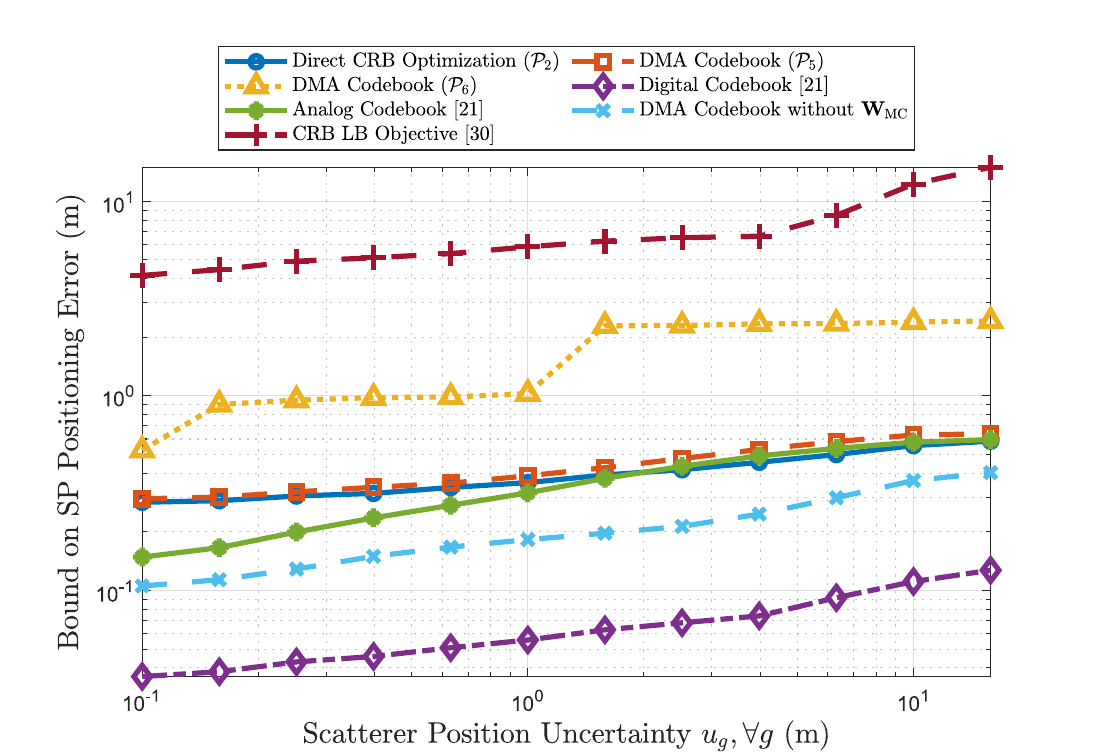}


  \includegraphics[width=0.5\textwidth]{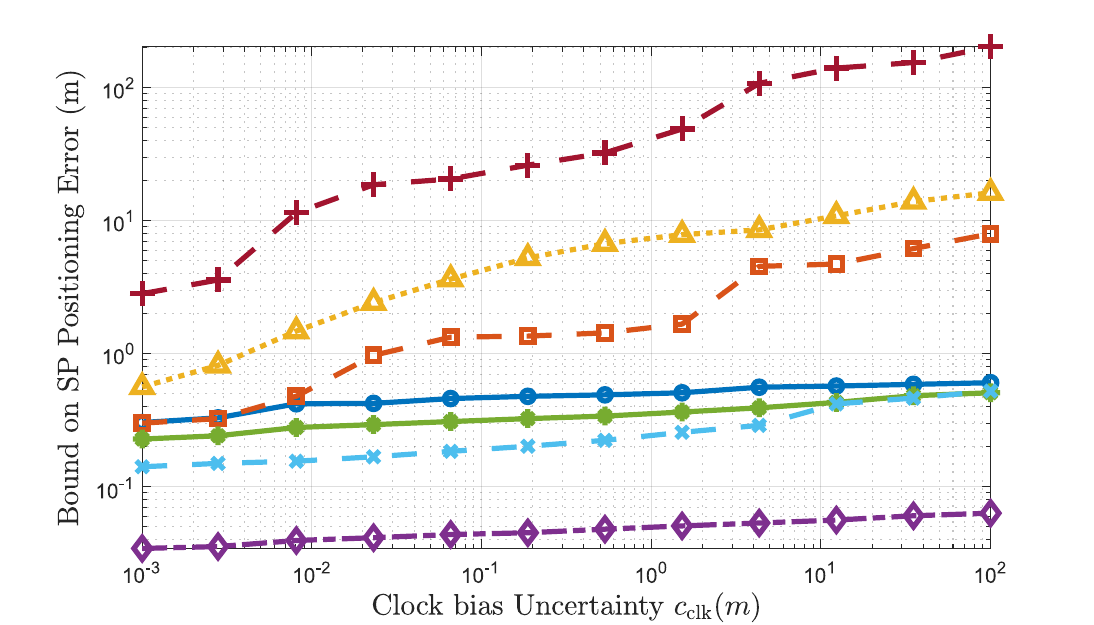}
  	\caption{\small{PEB versus clock bias uncertainty for the proposed mutual-coupling-aware BF strategies for Scenario 1 (\textit{top}) and Scenario 2 (\textit{bottom}), for $G=2$ SPs. In Scenario 1, $P=8$ points are needed to cover the entirety of the uncertainty region with $M=48$ beams, whereas, in Scenario 2, only $P = 2$ points with $M = 12$ beams are sufficient.}}
    \vspace{-0.4cm}
  \label{fig: Clk}
\end{figure}

The SPs' position estimation performance with all proposed DMA-based TX BF designs for bistatic sensing under varying clock bias uncertainties is illustrated in Fig.~\ref{fig: Clk}. In particular, this figure depicts SPs' PEB versus $\sigma_{\rm clk}$ in meters for Scenarios 1 (top) and~2 (bottom). Recall that all benchmark methods neglect mutual coupling across the DMA aperture. As expected, increasing $\sigma_{\rm clk}$ degrades positioning accuracy in both scenarios; this happens due to the added ambiguity along the time-of-arrival and range dimensions. It is shown that this degradation is more pronounced when the SPs' spatial uncertainty is larger (implying larger uncertainty region), i.e., in Scenario 1. However, it can be interestingly seen that the proposed mutual-coupling-aware BF designs achieve performance comparable to benchmarks. Notably, in cases with low position and clock uncertainties, the proposed codebook- and upper-bound-based BF designs perform on par with the direct CRB minimization approach, i.e., the proposed one performing alternating optimization of $\mathcal{P}_1$ and $\mathcal{P}_2$. On the other hand, as $\sigma_{\rm clk}$ increases, the PEB with codebook- and upper-bound-based designs deteriorates more rapidly than that with the direct CRB minimization approach. This behavior is attributed to mutual coupling that introduces nonlinear, spatially variant distortions in the DMA response, whose impact becomes increasingly detrimental under large $\sigma_{\rm clk}$ values.  

\begin{figure}[t!]
  \centering
  \includegraphics[width=0.5\textwidth]{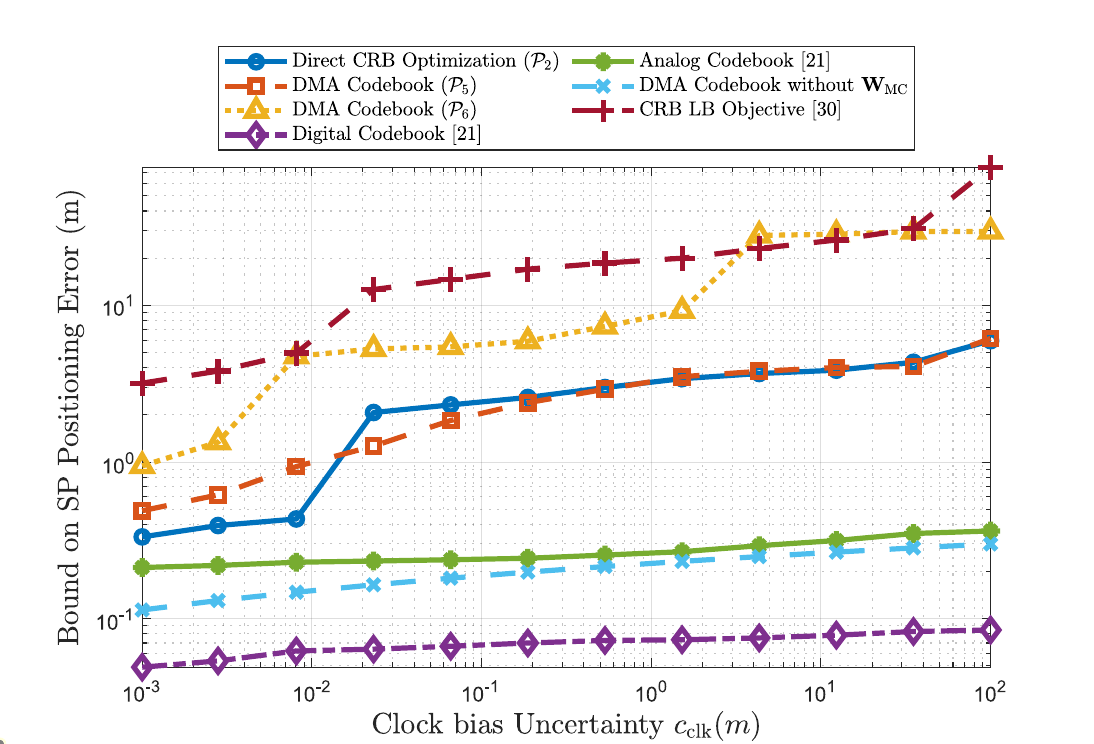}


  \includegraphics[width=0.5\textwidth]{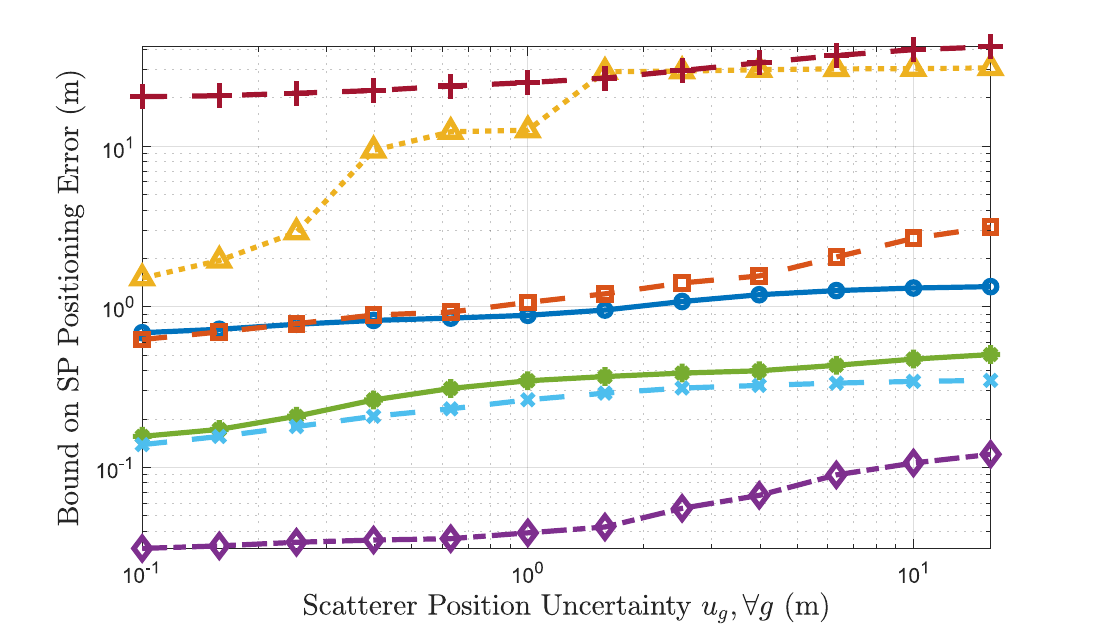}
  	\caption{\small{PEB versus SP position uncertainty for the proposed mutual-coupling-aware BF strategies for $\sigma_{\rm clk}=1$ meter (\textit{top}) and $\sigma_{\rm clk}=10^2$ meters (\textit{bottom}), for $G=2$ SPs. Notably, the number of frames $M$ for all the considered methods scales accordingly with the number of points needed to cover the uncertainty region in its entirety.}}
    \vspace{-0.4cm}
  \label{fig: SP_pos}
\end{figure}

Figure~\ref{fig: SP_pos} considers the same BF designs with Fig.~\ref{fig: Clk} and demonstrates the PEB as a function of the common radius $u_g$ in meters of all SPs' spherical uncertainty regions. The top row corresponds to a clock bias of $\sigma_{\rm clk} = 1$, while the bottom to $\sigma_{\rm clk} = 10^2$. Note that, as $u_g$ increases, the number of transmission frames $M$ scales accordingly to ensure full spatial coverage of the uncertainty region under a fixed transmit power constraint. It can be observed from the figure that, as expected, increasing $u_g$ leads to degraded localization performance. As seen, this degradation is more pronounced for larger $\sigma_{\rm clk}$. Nonetheless, it is evident that the proposed BF designs maintain competitive accuracy even under growing $u_g$, especially in scenarios with small $\sigma_{\rm clk}$ vaues. In contrast, under severe synchronization errors, the nonlinearities induced by mutual coupling with the considered EM-compliant DMA model result in PEB deterioration. It can be also seen that the proposed codebook- and upper-bound-based BF designs exhibit larger robustness to SPs' position uncertainty, outperforming their performance in Fig.~\ref{fig: Clk} under clock bias uncertainty at the RX side. Furthermore, it is depicted that, at small-to-moderate $u_g$ levels, the proposed designs yield comparable performance to their direct CRB minimization counterpart; however, a noticeable performance gap emerges as $u_g$ increases. This is again attributed to the fact that predesigned codebooks cannot fully adapt to the complex beam distortion patterns induced by mutual coupling and waveguide feed mapping over larger spatial regions. In contrast, the designed direct CRB minimization approach continuously reconfigures the BF weights to account for these distortions, leading to better spatial resolution and more accurate sensing.

\begin{figure*}[!t]

  \begin{subfigure}[t]{0.33\textwidth}
  \centering
    \includegraphics[width=\textwidth]{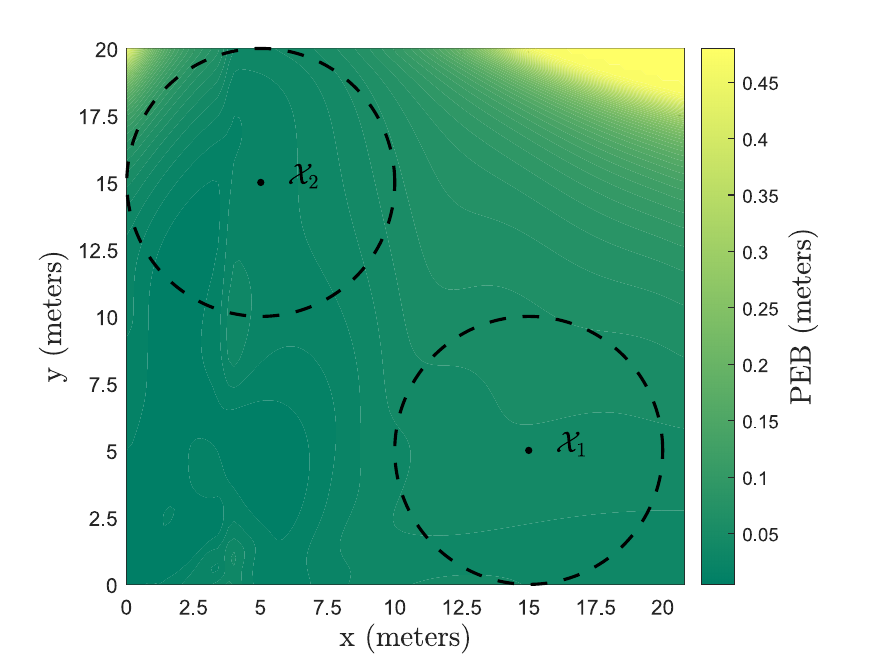}
  \end{subfigure}\hfill
  \begin{subfigure}[t]{0.33\textwidth}
  \centering
    \includegraphics[width=\textwidth]{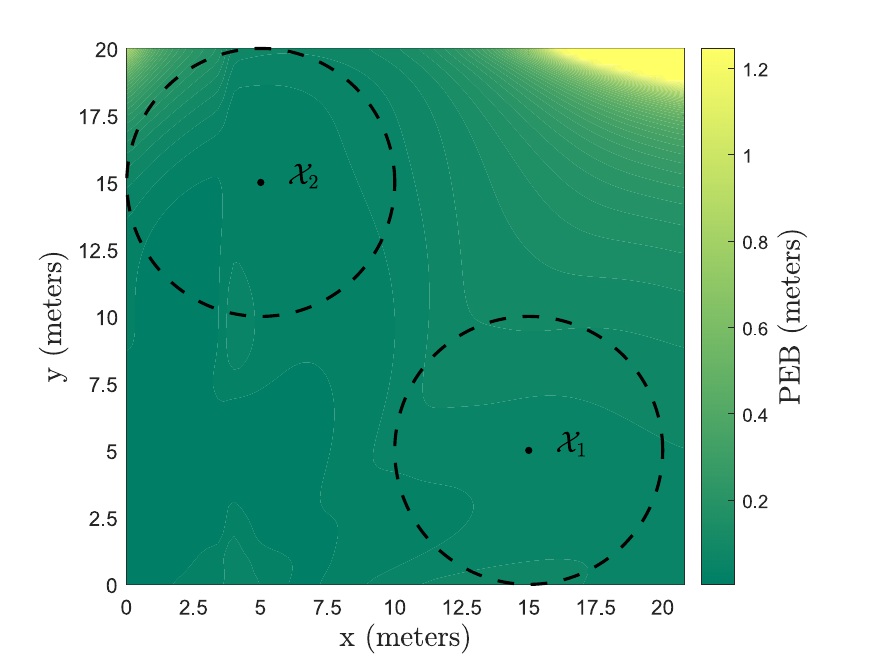}
  \end{subfigure}\hfill
  \begin{subfigure}[t]{0.33\textwidth}
  \centering
    \includegraphics[width=\textwidth]{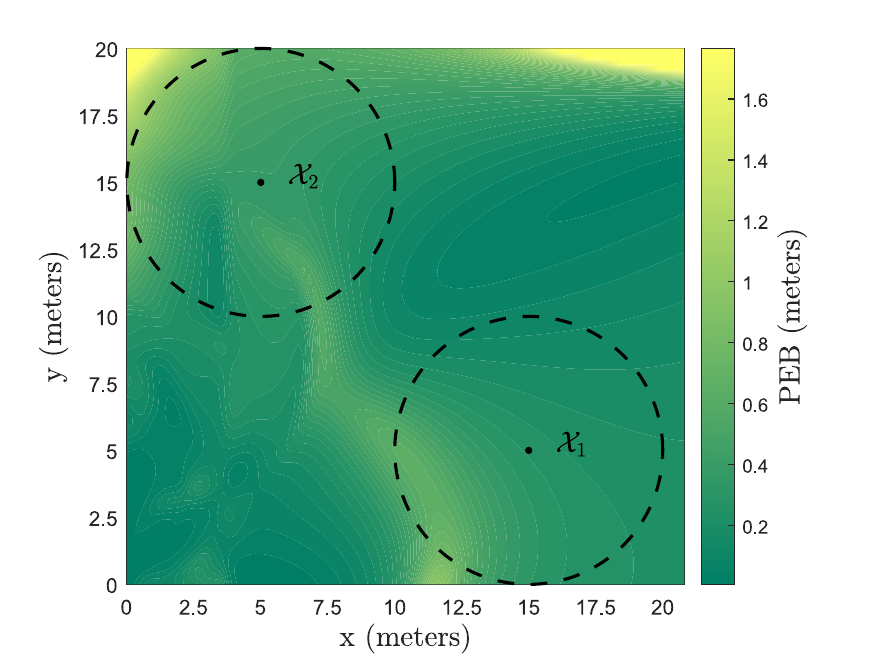}
  \end{subfigure}
    \begin{subfigure}[h]{0.33\textwidth}
  \centering
    \includegraphics[width=\textwidth]{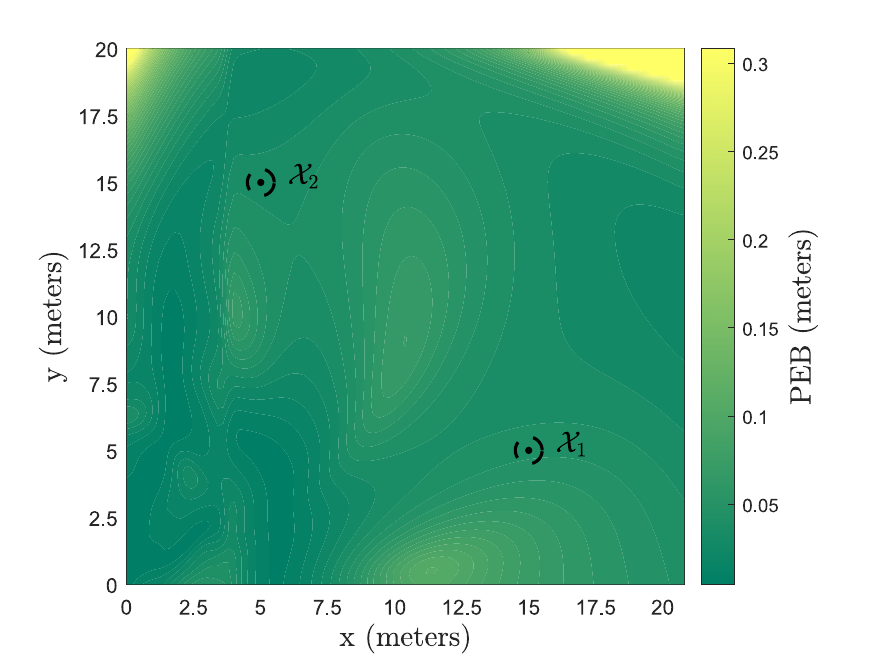}
  \caption{\small{Direct CRB minimization.}}
  \label{fig: Direct_CRB}
  \end{subfigure}
      \begin{subfigure}[h]{0.33\textwidth}
  \centering
    \includegraphics[width=\textwidth]{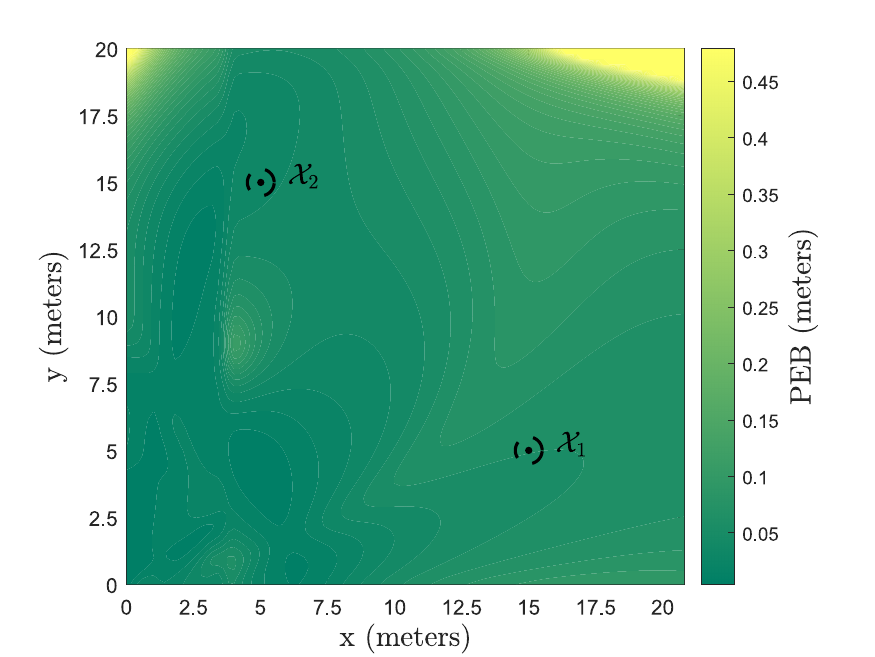}
    \caption{\small{BF strategy via Section~\ref{Sec: DMA_codebook_design}.}}
    \label{fig: PEB_DMA_code}
  \end{subfigure}
      \begin{subfigure}[h]{0.33\textwidth}
  \centering
    \includegraphics[width=\textwidth]{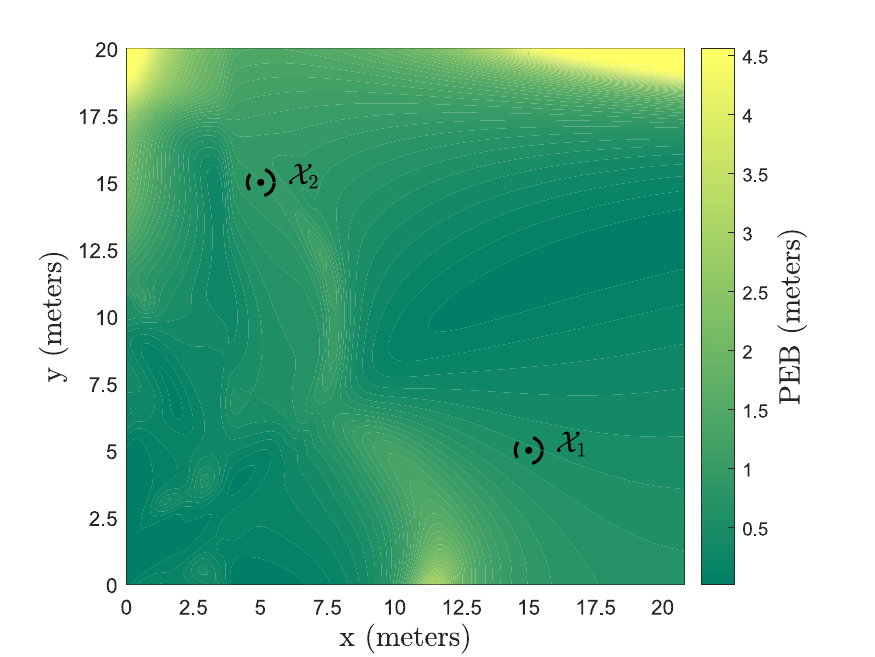}
    \caption{\small{BF strategy via Section~\ref{Sec: LB}.}}
    \label{fig: PEB_code_approx}
  \end{subfigure}
  \caption{\small{PEB across the $xy$-plane between the DMA-based TX and the RX of the considered bistatic sensing system using the following design approaches: \textit{i}) direct CRB minimization (solution of $\mathcal{P}_1$ and $\mathcal{P}_2$); \textit{ii}) Codebook-based BF strategy (Section~\ref{Sec: DMA_codebook_design}); and \textit{iii}) Codebook-based BF strategy via the CRB upper bound (Section~\ref{Sec: LB}). The \textit{top row} corresponds to Scenario 1, while the \textit{bottom row} corresponds to Scenario 2, when both evaluated with $\sigma_{\rm clk}=1$ meter. In all subfigures, the dotted circles represent the uncertainty regions $\mathcal{X}_1$ and $\mathcal{X}_2$ associated with the $G=2$ SPs.}}\vspace{-0.4cm}
  \label{fig: PEB_heatmap}
\end{figure*}

By focusing on the spatial region between the DMA-based TX and the RX, Fig.~\ref{fig: PEB_heatmap} depicts the PEB heatmaps with both proposed codebook- and upper-bound-based BF designs and the one devised via the direct CRB minimization, considering both SPs' position uncertainty Scenarios 1 (top row) and~2 (bottom row) as well as $\sigma_{\rm clk}=1$~m. The dotted circles represent the uncertainty regions $\mathcal{X}_1$ and $\mathcal{X}_2$ associated with the $G=2$ SPs in both scenarios. A consistent PEB performance throughout both the entire uncertainty regions can be observed for both scenarios when the position uncertainty is small. As expected, the direct CRB minimization approach achieves the best robust sensing performance; this is also showcased in Figs.~\ref{fig: Clk} and~\ref{fig: SP_pos}. Interestingly, both the codebook- and upper-bound-based designs deliver comparable results, thus, validating their effectiveness as practical alternatives with reasonable computational complexity. It is also worth noting some additional important nuances: \textit{i}) it is evident that an increase in position uncertainty leads to wider sidelobes emanating from the RX, a fact that degrades the resolution of angular estimation; \textit{ii}) along the TX's boresight angle, range ambiguity becomes more pronounced, causing noticeable spikes in the PEB performance; and \textit{iii}) the smoother PEB contours achieved by the direct CRB minimization approach resulted from the continuous optimization over the full beam space. On the other hand, the codebook- and upper-bound-based designs rely on discrete, hardware-constrained beams that inevitably introduce spatial quantization effects.

\begin{figure*}[!t]
  \begin{subfigure}[t]{0.33\textwidth}
  \centering
    \includegraphics[width=\textwidth]{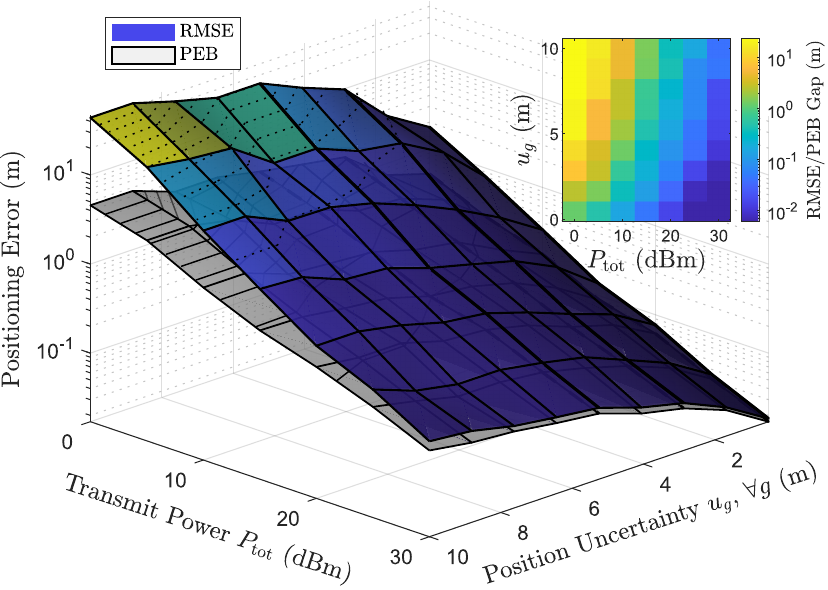}
    \caption{\small{Direct CRB minimization.}}
  \end{subfigure}\hfill
  \begin{subfigure}[t]{0.33\textwidth}
  \centering
    \includegraphics[width=\textwidth]{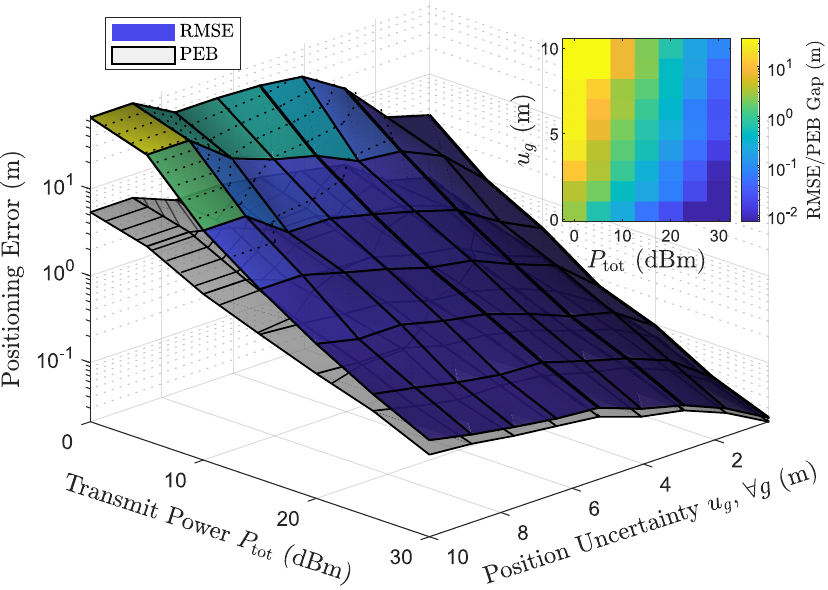}
    \caption{\small{BF strategy via Section~\ref{Sec: DMA_codebook_design}.}}
  \end{subfigure}\hfill
  \begin{subfigure}[t]{0.33\textwidth}
  \centering
    \includegraphics[width=\textwidth]{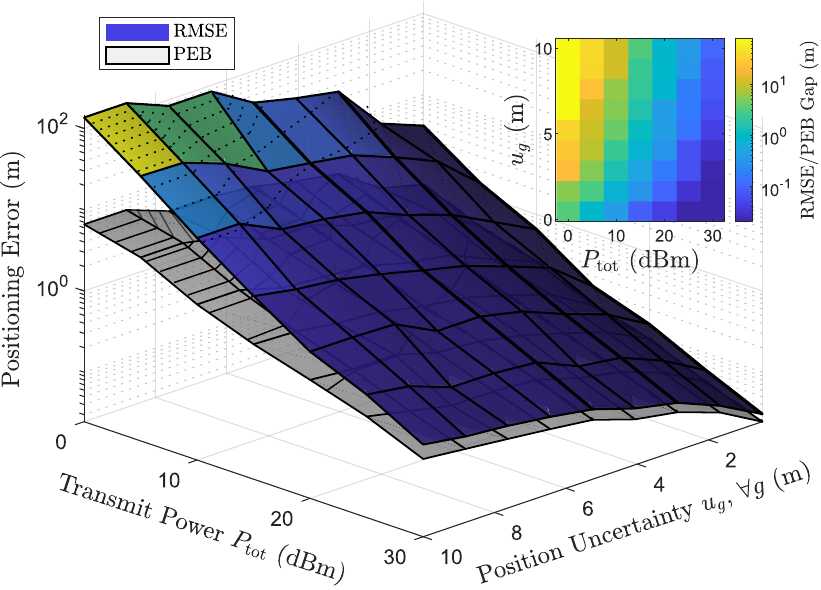}
    \caption{\small{BF strategy via Section~\ref{Sec: LB}.}}
  \end{subfigure}
\caption{\small{RMSE of SPs position and corresponding PEB (meters) as functions of the total transmit power \(P_{\rm tot}\) (dBm) and the position-uncertainty radius \(u_g\) (meters) for both $G=2$ SPs. Three TX BF design strategies are compared: \textit{i}) direct CRB minimization via solving \(\mathcal{P}_1\) and \(\mathcal{P}_2\); \textit{ii}) the proposed codebook-based BF design in Section~\ref{Sec: DMA_codebook_design}; and \textit{iii}) the codebook-based BF design relying on the CRB upper bound in Section~\ref{Sec: LB}. The clock bias uncertainty was set as \(\sigma_{\rm clk}=1\)~m and Algorithm~\ref{Algo} was utilized for the SPs position estimation. The inset depicts the RMSE--PEB gap heatmap (i.e., RMSE minus PEB), highlighting the tightness of the bound across operating points.}}\vspace{-0.4cm}
  \label{fig: MLE}
\end{figure*}

Figure~\ref{fig: MLE} assesses the proposed bistatic sensing design in Algorithm~\ref{Algo} via the SPs position RMSE together with the corresponding PEB versus the total TX power \(P_{\mathrm{tot}}\) and the common uncertainty radius $u_g$ for both $G=2$ SPs. We have fixed \(\sigma_{\mathrm{clk}}=1\)~m, \(L_{\max}=50\), and \(\kappa=4\), and compared the proposed designs as in Fig.~\ref{fig: PEB_heatmap}. As expected, increasing \(P_{\mathrm{tot}}\) improves both RMSE and PEB, whereas increasing \(u_g\) degrades performance, since robustness over a larger uncertainty set spreads transmit energy across more spatial hypotheses, weakening focusing and thus localization resolution, similar to Fig.~\ref{fig: SP_pos}. Among the BF designs, direct CRB minimization achieves the lowest errors, with gains that grow with \(u_g\), and continuous weight optimization better accommodates the EM-compliant DMA response mitigating spatially varying distortions. Codebook-based schemes remain competitive for small-to-moderate \(u_g\), but their gap widens at larger \(u_g\) due to the limited ability of predesigned, hardware-constrained beams to maintain uniformly sharp focusing under coupling/feed-induced beam perturbations. The inset RMSE-PEB gap heatmaps indicate tightness of the objective: the proposed low complexity estimator closely tracks the bound as \(P_{\rm tot}\) increases, with a small gap even for larger uncertainty.

\section{Conclusions and Future Directions}\label{Sec: Conclusion}
In this paper, we presented an EM-compliant BF design framework for robust bistatic sensing with a DMA-based TX, addressing a key gap in the scientific literature that often overlooks the physical and structural constraints of this emerging XL antenna array architecture. By incorporating mutual coupling effects into our system modeling, we first introduced a tractable approximation of the DMA response that enables its efficient BF optimization. Building on this approximation, we formulated a robust optimization framework to minimize the worst-case PEB performance under spatial SP uncertainties and TX-RX synchronization inconsistencies. To tackle the complexity overhead induced by the numerous optimization variables and LMI constraints, we devised two low complexity alternatives: a DMA-compatible codebook-based BF design and another based on a CRB lower bound approximation. In addition, a multi-target parameter estimation
approach, consistent with the proposed PEB framework and performing region-aware sequential position initialization followed by structured synchronization parameter search and an alternating refinement stage, is presented. 
 Through extensive simulations, we demonstrated that accurately modeling mutual coupling is critical for achieving reliable sensing performance, particularly under high positional and synchronization uncertainties. Our results also showcased that waveguide feed mapping, combined with the structural constraints of DMAs, limit their ability to closely replicate the performance of fully digital BF architectures. However, the proposed EM-compliant architectures were shown to offer comparable performance to fully digital and analog benchmarks, especially under moderate uncertainty, and the proposed localization scheme closely tracked the derived CRB even in high-uncertainty scenarios. Overall, this paper highlighted the feasibility and performance potential of realistic DMA-based bistatic sensing, offering a promising direction for future wireless sensing applications. Future direction include mutual-coupling-aware 2D waveguide-fed metasurface antennas with extensions to XL configurations, scalable frequency selective hybrid analog and digital BF solutions for XL apertures, as well as joint sensing and communications applications. 

\appendices
\section{}\label{App}
Let us assume that $\W_{\rm TX}$ and $\W\triangleq\W_{\rm RX}\W_{\rm RX}^{\rm H}$ are fixed. Then, each $(i,j)$-th element ($i,j=1,\ldots,7G$) of the FIM $\J$ in \eqref{eq: FIM} depends linearly on $\X\triangleq\F\F^{\rm H}$ as shown below:
\begin{align}\label{eq: FIM_lin_X}
    \nonumber&[\J]_{i,j}=\frac{2T}{\sigma^2}\sum_{m,k}\Re\left\{\f_m^{\rm H}\W_{\rm TX}^{\rm H}\frac{\partial\H_k^{\rm H}}{\partial\xi_i}\W\frac{\partial\H_k}{\partial\xi_j}\W_{\rm TX}\f_m\right\}\!
    \\\nonumber&=\frac{2T}{\sigma^2}\sum_{m,k}\Re\left\{{\rm Tr}\left(\f_m\f_m^{\rm H}\W_{\rm TX}^{\rm H}\frac{\partial\H_k^{\rm H}}{\partial\xi_i}\W\frac{\partial\H_k}{\partial\xi_j}\W_{\rm TX}\right)\right\}\!
    \\&=\frac{2T}{\sigma^2}\sum_{k=1}^{K}\Re\left\{{\rm Tr}\left(\X\W_{\rm TX}^{\rm H}\frac{\partial\H_k^{\rm H}}{\partial\xi_i}\W\frac{\partial\H_k}{\partial\xi_j}\W_{\rm TX}\right)\right\}\!. \tag{A-1}
\end{align}

Similarly, when $\X$ and $\W$ are fixed and $\W_{\rm TX}$ is modeled according to \eqref{eq: sec_order}, we can exploit the affine dependence of $\W_{\rm TX}$ on the real-valued tunable reactive parameters $\c$. In particular, under the second order approximation in \eqref{eq: sec_order} we can reformulate $\W_{\rm TX}$ as $\W_{\rm TX} = \W+\sum_{n=1}^{N_{\rm T}}c_n\W_n$, where $\W_0=(\W_{\rm MC}^{-1}+\jmath\W_{\rm MC}^{-1}{\rm diag}(\boldsymbol{\nu})\W_{\rm MC}^{-1})\P_{\rm SA}$ and $\W_n=-\jmath\W_{\rm MC}^{-1}\e_n\e_n^{\rm T}\W_{\rm MC}^{-1}\P_{\rm SA}$ with $\e_n$ being the $n$-th column of the $N_{\rm T}\times N_{\rm T}$ identity matrix. Continuing, the product of the analog and digital BF parameters for the $m$-th frame can be rewritten as $\W_{\rm TX}\f_m=\boldsymbol{\Omega}_m\bar{\c}$, where $\boldsymbol{\Omega}_m\triangleq[\W_1\f_m,\ldots,\W_{N_{\rm T}}\f_m, \W_0\f_m]$ and $\bar{\c}\triangleq[\c^{\rm T},1]^{\rm T}$. Thus, starting from the FIM expression as in \eqref{eq: FIM_lin_X}, we can rewrite the FIM with respect to the real positive semidefinite matrix $\C\triangleq\bar{\c}\bar{\c}^{\rm H}$, as follows:
\begin{align}\label{eq: FIM_lin_C}
    \nonumber&[\J]_{i,j}=\frac{2T}{\sigma^2}\sum_{m,k}\Re\left\{\bar{\c}^{\rm H}\boldsymbol{\Omega}_m^{\rm H}\frac{\partial\H_k^{\rm H}}{\partial\xi_i}\W\frac{\partial\H_k}{\partial\xi_j}\boldsymbol{\Omega}_m\bar{\c}\right\}\!
    \\\nonumber&=\frac{2T}{\sigma^2}\sum_{m,k}\Re\left\{{\rm Tr}\left(\C\boldsymbol{\Omega}_m^{\rm H}\frac{\partial\H_k^{\rm H}}{\partial\xi_i}\W\frac{\partial\H_k}{\partial\xi_j}\boldsymbol{\Omega}_m\right)\right\}. \tag{A-2}
\end{align}

\section{}\label{AppB}
Let us consider the problem formulation $\mathcal{P}$ under the assumption that each $\mathcal{X}_g$ consists of a single point corresponding to the true target position $\p_g$, rather than a continuous set of candidate locations. In this ideal case, unlike in $\mathcal{P}_1$ and $\mathcal{P}_2$, there is no need for discretization over $P$ uniformly spaced points. To this end, following a reasoning analogous to~\cite[Proposition 1]{fascista2022ris},\cite{li2007range} and assuming perfect knowledge of the 3D AoD parameters $(\theta_{{\rm a},g}, \theta_{{\rm e},g})$ $\forall g = 1, \dots, G$ as well as the RX position, the joint analog and digital covariance matrix in the PEB expression can be written similar to Appendix~\ref{App} as $\Z \triangleq \W_{\rm TX}\X\W_{\rm TX}^{\rm H}$. It then follows that the dependence of $\J$ on $\Z$ arises solely through the matrices $\A_{\rm TX}, \dot{\A}_{\theta_{\rm a}}$, and $\dot{\A}_{\theta_{\rm e}}$, which are defined as follows:
\begin{align*}
&\mathbf{A}_{\rm TX} \triangleq \left[\mathbf{a}_{\rm TX}(\theta_{{\rm a},1}, \theta_{{\rm e},1}), \dots, \mathbf{a}_{\rm TX}(\theta_{{\rm a},G}, \theta_{{\rm e},G})\right], \\
&\dot{\mathbf{A}}_{\rm\theta_{\rm a}} \triangleq \left[\frac{\partial \mathbf{a}_{\rm TX}(\theta_{{\rm a},1}, \theta_{{\rm e},1})}{\partial \theta_{{\rm a},1}}, \dots, \frac{\partial \mathbf{a}_{\rm TX}(\theta_{{\rm a},G}, \theta_{{\rm e},G})}{\partial \theta_{{\rm a},G}}\right], \\
&\dot{\mathbf{A}}_{\rm\theta_{\rm e}} \triangleq \left[\frac{\partial \mathbf{a}_{\rm TX}(\theta_{{\rm a},1}, \theta_{{\rm e},1})}{\partial \theta_{{\rm e},1}}, \dots, \frac{\partial \mathbf{a}_{\rm TX}(\theta_{{\rm a},G}, \theta_{{\rm e},G})}{\partial \theta_{{\rm e},G}}\right].
\end{align*}
Thus, as shown in \cite{li2007range}, the optimal solution of $\mathcal{P}$ for this single-point case admits the form $\Z_{\rm opt} = \U_{\rm TX} \boldsymbol{\Lambda} \U_{\rm TX}^{\rm H}$, where $\boldsymbol{\Lambda} \in \mathbb{C}^{3G \times 3G}$ is a diagonal, positive semidefinite matrix and $\U_{\rm TX} \triangleq [ \A_{\rm TX}, \dot{\A}_{\theta_{\rm a}}, \dot{\A}_{\theta_{\rm e}} ]$, i.e., the column space of the optimal covariance matrix $\Z_{\rm opt}$ is spanned by the TX steering vectors pointing at the directions of the $G$ targets as well as their first-order angular derivatives. Note that the optimal covariance structure can be derived only in the case where PEB is minimized for a single known parameter vector $\widetilde{\boldsymbol{\xi}}$, and not for the worst-case PEB over an uncertainty region, as considered in the problem formulation $\mathcal{P}$. Leveraging the structure of the optimal solution to the single-point case for $\mathcal{P}$, we can design the hybrid digital and analog (and mutual-coupling-aware) DMA codebook, as detailed in Section~\ref{Sec: DMA_codebook_design}.

\bibliographystyle{IEEEtran}
\bibliography{ms}
\end{document}

%% file: Definitions.tex
\renewcommand{\a}{\mathbf{a}}

\renewcommand{\c}{\mathbf{c}}
\renewcommand{\d}{\mathbf{d}}
\newcommand{\e}{\mathbf{e}}
\newcommand{\f}{\mathbf{f}}
\newcommand{\g}{\mathbf{g}}

\renewcommand{\k}{\mathbf{k}}

\newcommand{\n}{\mathbf{n}}

\newcommand{\p}{\mathbf{p}}

\renewcommand{\r}{\mathbf{r}}

\renewcommand{\u}{\mathbf{u}}

\newcommand{\x}{\mathbf{x}}
\newcommand{\y}{\mathbf{y}}

\newcommand{\A}{\mathbf{A}}
\newcommand{\B}{\mathbf{B}}
\newcommand{\C}{\mathbf{C}}

\newcommand{\F}{\mathbf{F}}
\newcommand{\G}{\mathbf{G}}
\renewcommand{\H}{\mathbf{H}}
\newcommand{\I}{\mathbf{I}}
\newcommand{\J}{\mathbf{J}}
\newcommand{\K}{\mathbf{K}}

\renewcommand{\P}{\mathbf{P}}
\newcommand{\Q}{\mathbf{Q}}
\newcommand{\R}{\mathbf{R}}
\renewcommand{\S}{\mathbf{S}}
\newcommand{\T}{\mathbf{T}}
\newcommand{\U}{\mathbf{U}}
\newcommand{\V}{\mathbf{V}}
\newcommand{\W}{\mathbf{W}}
\newcommand{\X}{\mathbf{X}}

\newcommand{\Z}{\mathbf{Z}}

\newcommand{\Compl}{\mbox{$\mathbb{C}$}}

\newcommand{\argmin}{\operatornamewithlimits{argmin}}
\newcommand{\argmax}{\operatornamewithlimits{argmax}}

\renewcommand{\Im}{\mathrm{Im}}

\renewcommand{\Re}{\mathrm{Re}}